\documentclass[aip,amsmath,amssymb,preprint,a4paper,floatfix,nofootinbib]{revtex4-1}

\usepackage{graphicx}
\usepackage{subfigure}
\usepackage[usenames,dvipsnames,table]{xcolor}
\usepackage{color}
\usepackage{relsize}

\begin{document}

\title{First light demonstration of the integrated superconducting spectrometer}

\author{Akira~Endo}
    \email[]{a.endo@tudelft.nl}
    \affiliation{Faculty of Electrical Engineering, Mathematics and Computer Science, Delft University of Technology, Mekelweg 4, 2628 CD Delft, the Netherlands.}
    \affiliation{Kavli Institute of NanoScience, Faculty of Applied Sciences, Delft University of Technology, Lorentzweg 1, 2628 CJ Delft, The Netherlands.}
    
\author{Kenichi~Karatsu}
    \affiliation{SRON---Netherlands Institute for Space Research, Sorbonnelaan 2, 3584 CA Utrecht, The Netherlands.}
    \affiliation{Faculty of Electrical Engineering, Mathematics and Computer Science, Delft University of Technology, Mekelweg 4, 2628 CD Delft, the Netherlands.}

\author{Yoichi~Tamura}
    \affiliation{Division of Particle and Astrophysical Science, Graduate School of Science, Nagoya University, Aichi 464-8602, Japan.}

\author{Tai~Oshima}
    \affiliation{National Astronomical Observatory of Japan, Mitaka, Tokyo 181-8588, Japan.}
    \affiliation{The Graduate University for Advanced Studies (SOKENDAI), 2-21-1 Osawa, Mitaka, Tokyo 181-0015, Japan}
    
\author{Akio~Taniguchi}
    \affiliation{Division of Particle and Astrophysical Science, Graduate School of Science, Nagoya University, Aichi 464-8602, Japan.}
    
\author{Tatsuya~Takekoshi}
    \affiliation{Institute of Astronomy, Graduate School of Science, The University of Tokyo, 2-21-1 Osawa, Mitaka, Tokyo 181-0015, Japan.}
    \affiliation{Graduate School of Informatics and Engineering, The University of Electro-Communications, Cho-fu, Tokyo 182-8585, Japan}

\author{Shin'ichiro~Asayama}
    \affiliation{National Astronomical Observatory of Japan, Mitaka, Tokyo 181-8588, Japan.}

\author{Tom~J.~L.~C.~Bakx}
    \affiliation{Division of Particle and Astrophysical Science, Graduate School of Science, Nagoya University, Aichi 464-8602, Japan.}
    \affiliation{National Astronomical Observatory of Japan, Mitaka, Tokyo 181-8588, Japan.}
    \affiliation{School of Physics \& Astronomy, Cardiff University, The Parade, Cardiff, CF24 3AA UK}

\author{Sjoerd~Bosma}
    \affiliation{Faculty of Electrical Engineering, Mathematics and Computer Science, Delft University of Technology, Mekelweg 4, 2628 CD Delft, the Netherlands.}
    
\author{Juan~Bueno}
    \affiliation{SRON---Netherlands Institute for Space Research, Sorbonnelaan 2, 3584 CA Utrecht, The Netherlands.}

\author{Kah~Wuy~Chin}
    \affiliation{National Astronomical Observatory of Japan, Mitaka, Tokyo 181-8588, Japan.}
    \affiliation{Department of Astronomy, School of Science, University of Tokyo, Bunkyo, Tokyo, 113-0033, Japan}
    
\author{Yasunori~Fujii}
    \affiliation{National Astronomical Observatory of Japan, Mitaka, Tokyo 181-8588, Japan.}
    
\author{Kazuyuki~Fujita}
    \affiliation{Institute of Low Temperature Science, Hokkaido University, Sapporo 060-0819, Japan}

\author{Robert~Huiting}
    \affiliation{SRON---Netherlands Institute for Space Research, Sorbonnelaan 2, 3584 CA Utrecht, The Netherlands.}

\author{Soh~Ikarashi}
    \affiliation{Faculty of Electrical Engineering, Mathematics and Computer Science, Delft University of Technology, Mekelweg 4, 2628 CD Delft, the Netherlands.}

\author{Tsuyoshi~Ishida}
    \affiliation{Institute of Astronomy, Graduate School of Science, The University of Tokyo, 2-21-1 Osawa, Mitaka, Tokyo 181-0015, Japan.}

\author{Shun~Ishii}
    \affiliation{National Astronomical Observatory of Japan, Mitaka, Tokyo 181-8588, Japan.}
    \affiliation{Joint ALMA Observatory, Alonso de C\'ordova 3107, Vitacura, Santiago, Chile.}

\author{Ryohei~Kawabe}
    \affiliation{National Astronomical Observatory of Japan, Mitaka, Tokyo 181-8588, Japan.}
    \affiliation{The Graduate University for Advanced Studies (SOKENDAI), 2-21-1 Osawa, Mitaka, Tokyo 181-0015, Japan}
    \affiliation{Department of Astronomy, School of Science, University of Tokyo, Bunkyo, Tokyo, 113-0033, Japan}

\author{Teun~M.~Klapwijk}
    \affiliation{Kavli Institute of NanoScience, Faculty of Applied Sciences, Delft University of Technology, Lorentzweg 1, 2628 CJ Delft, The Netherlands.}
    \affiliation{Physics Department, Moscow State Pedagogical University, 119991 Moscow, Russia.}
    
\author{Kotaro~Kohno}
    \affiliation{Institute of Astronomy, Graduate School of Science, The University of Tokyo, 2-21-1 Osawa, Mitaka, Tokyo 181-0015, Japan.}
    \affiliation{Research Center for the Early Universe, Graduate School of Science, The University of Tokyo, 7-3-1 Hongo, Bunkyo-ku, Tokyo 113-0033, Japan}

\author{Akira~Kouchi}
    \affiliation{Institute of Low Temperature Science, Hokkaido University, Sapporo 060-0819, Japan}

\author{Nuria~Llombart}
    \affiliation{Faculty of Electrical Engineering, Mathematics and Computer Science, Delft University of Technology, Mekelweg 4, 2628 CD Delft, the Netherlands.}

\author{Jun~Maekawa}
    \affiliation{National Astronomical Observatory of Japan, Mitaka, Tokyo 181-8588, Japan.}

\author{Vignesh~Murugesan}
    \affiliation{SRON---Netherlands Institute for Space Research, Sorbonnelaan 2, 3584 CA Utrecht, The Netherlands.}

\author{Shunichi~Nakatsubo}
    \affiliation{Institute of Space and Astronautical Science, Japan Aerospace Exploration Agency, Sagamihara 252-5210, Japan.}

\author{Masato~Naruse}
    \affiliation{Graduate School of Science and Engineering, Saitama University, 255, Shimo-okubo, Sakura, Saitama 338-8570, Japan.}
    
\author{Kazushige~Ohtawara}
    \affiliation{National Astronomical Observatory of Japan, Mitaka, Tokyo 181-8588, Japan.}
    
\author{Alejandro~Pascual~Laguna}
    \affiliation{SRON---Netherlands Institute for Space Research, Sorbonnelaan 2, 3584 CA Utrecht, The Netherlands.}
    \affiliation{Faculty of Electrical Engineering, Mathematics and Computer Science, Delft University of Technology, Mekelweg 4, 2628 CD Delft, the Netherlands.}

\author{Junya~Suzuki}
    \affiliation{High Energy Accelerator Research Organization (KEK), 1-1 Oho, Tsukuba, Ibaraki, 305-0801, Japan.}

\author{Koyo~Suzuki}
    \affiliation{Division of Particle and Astrophysical Science, Graduate School of Science, Nagoya University, Aichi 464-8602, Japan.}
    
\author{David~J.~Thoen}
    \affiliation{Faculty of Electrical Engineering, Mathematics and Computer Science, Delft University of Technology, Mekelweg 4, 2628 CD Delft, the Netherlands.}
    \affiliation{Kavli Institute of NanoScience, Faculty of Applied Sciences, Delft University of Technology, Lorentzweg 1, 2628 CJ Delft, The Netherlands.}

\author{Takashi~Tsukagoshi}
    \affiliation{National Astronomical Observatory of Japan, Mitaka, Tokyo 181-8588, Japan.}

\author{Tetsutaro~Ueda}
    \affiliation{Division of Particle and Astrophysical Science, Graduate School of Science, Nagoya University, Aichi 464-8602, Japan.}

\author{Pieter~J.~de~Visser}
    \affiliation{SRON---Netherlands Institute for Space Research, Sorbonnelaan 2, 3584 CA Utrecht, The Netherlands.}

\author{Paul~P.~van~der~Werf}
    \affiliation{ 
Leiden Observatory, Leiden University, PO Box 9513, NL-2300 RA Leiden, The Netherlands.
    }

\author{Stephen~J.~C.~Yates}
    \affiliation{SRON---Netherlands Institute for Space Research,  Landleven 12, 9747 AD Groningen, The Netherlands.}

\author{Yuki~Yoshimura}
    \affiliation{Institute of Astronomy, Graduate School of Science, The University of Tokyo, 2-21-1 Osawa, Mitaka, Tokyo 181-0015, Japan.}

\author{Ozan~Yurduseven}
    \affiliation{Faculty of Electrical Engineering, Mathematics and Computer Science, Delft University of Technology, Mekelweg 4, 2628 CD Delft, the Netherlands.}

\author{Jochem~J.~A.~Baselmans}
    \affiliation{SRON---Netherlands Institute for Space Research, Sorbonnelaan 2, 3584 CA Utrecht, The Netherlands.}
    \affiliation{Faculty of Electrical Engineering, Mathematics and Computer Science, Delft University of Technology, Mekelweg 4, 2628 CD Delft, the Netherlands.}

\maketitle  

\begin{quotation}
Ultra-wideband 3D imaging spectrometry in the millimeter-submillimeter (mm-submm) band is an essential tool for uncovering the dust-enshrouded portion of the cosmic history of star formation and galaxy evolution\cite{2019arXiv190304779G,2017arXiv170902389F,2016SPIE.9906E..26K}. However, it is challenging to scale up conventional coherent heterodyne receivers\cite{2007ASPC..375...71E} or free-space diffraction techniques\cite{2011ITTST...1..241S} to sufficient bandwidths ($\geq$1 octave) and numbers of spatial pixels\cite{2017arXiv170902389F, 2016SPIE.9906E..26K} ($>$$10^2$). Here we present the design and first astronomical spectra of an intrinsically scalable, integrated superconducting spectrometer\cite{E1}, which covers 332--377~GHz with a spectral resolution of $F/\Delta F \sim 380$. It combines the multiplexing advantage of microwave kinetic inductance detectors (MKIDs\cite{2003Natur.425..817D}) with planar superconducting filters for dispersing the signal in a single, small superconducting integrated circuit. We demonstrate the two key applications for an instrument of this type: as an efficient redshift machine, and as a fast multi-line spectral mapper of extended areas. 
The line detection sensitivity is in excellent agreement with the instrument design and laboratory performance, reaching the atmospheric foreground photon noise limit on sky. The design can be scaled to bandwidths in excess of an octave, spectral resolution up to a few thousand and frequencies up to $\sim$1.1 THz. The miniature chip footprint of a few $\mathrm{cm^2}$ allows for compact  multi-pixel spectral imagers, which would enable spectroscopic direct imaging and large volume spectroscopic surveys that are several orders of magnitude faster than what is currently possible\cite{2019arXiv190304779G,2017arXiv170902389F,2016SPIE.9906E..26K}.
\end{quotation}

Galaxies grow through mergers and by drawing gas from their environment, while internally forming new stars and feeding matter onto their central supermassive black hole\cite{2018Sci...362.1034D}. These evolutionary processes occurred in a decadal redshift range of $1+z \sim1\text{--}10$. Hence, observationally studying a significant fraction of this history requires a very broad spectral bandwidth of a few octaves. The most violent phases of star-formation occur in thick clouds of dust, which absorbs the ultraviolet to optical light and reradiates this light in the far-infrared to mm-submm wavelength range, giving rise to optically-faint submm-bright galaxies (SMGs)\cite{Casey:2014gr}. However, spectroscopic redshift measurements and subsequent studies of the spectral lines 
from these SMGs are severely limited by the narrow bandwidth 
(up to $\sim$36 GHz\cite{2007ASPC..375...71E}) and small number of spatial pixels (exceptionally up to 64 pixels\cite{2010SPIE.7741E..0XG}) of conventional coherent spectrometers, which require multiple tunings and long exposure times. Quasioptical spectrometers have shown wide-band performance\cite{2011ITTST...1..241S}, but dispersive optical elements for the mm-submm band are large, making it difficult to scale this type of spectrometers to many spatial pixels. 

The integrated superconducting spectrometer (ISS, hereafter)\cite{2012JLTP..167..341E,Shirokoff:2012fx, Cataldo2018} is an instrument concept that was invented exactly to fill the gap between imaging and high resolution spectroscopy. The key concept of the ISS is to perform spectroscopy in a superconducting circuit fabricated on a small chip of a few cm$^2$ in area, using an array of bandpass filters as the dispersive element analogous to a classical filterbank for lower microwave frequencies. The main advantage of an ISS (or grating spectrometer) over a Fourier transform spectrometer (FTS) is that it is a dispersive spectrometer, which reduces the detection bandwidth and hence photon noise contribution, giving an observing speed improvement\cite{2014SPIE.9150E..0JS}. The ISS instantaneous bandwidth is limited by the antenna bandwidth and the filter design, which allows 1:2 or even 1:3 
bandwidth\cite{2012JLTP..167..341E,Shirokoff:2012fx, Cataldo2018,2017ApPhL.110w3503B,OBrient:2013hc}. Many spectrometers could be integrated on a single wafer, allowing for monolithic spectroscopic-imaging focal-plane architectures. Because the detectors are incoherent (i.e., they measure only the power and not the phase), the sensitivity of the ISS is not subject to quantum noise, giving ISSs a fundamental sensitivity advantage over heterodyne receivers\cite{2011ITTST...1..241S} when operated in low-foreground/background conditions typical for a space observatory\cite{2018NatAs...2..596B,2014NatCo...5E3130D}. Key technological ingredients of the ISS have been demonstrated in the laboratory, including the filterbank\cite{Endo:2013ky,Wheeler:2016dr}, antenna coupling\cite{Wheeler:2016dr,OBrient:2013hc}, and detection of emission lines from a gas cell\cite{E1}.

In this Letter, we present the first astronomical spectra obtained with an ISS, from the on-sky test of DESHIMA\cite{E1} (Deep Spectroscopic High-redshift Mapper) on the ASTE (Atacama Submillimeter Telescope Experiment) 10 m telescope\cite{2004SPIE.5489..763E}. DESHIMA instantaneously observes the 332--377 GHz band in fractional frequency steps of $F/\Delta F\sim 380$, matched to the $\sim$330--365 GHz atmospheric window (see Fig.~2h,i).  The instrument sensitivity is photon-noise limited, reaching a noise equivalent power (NEP) of $\sim$$3\times 10^{-16}\ \mathrm{W\ Hz^{-0.5}}$ under optical loading power levels representative of observing conditions on a ground-based submm telescope. The detailed design and laboratory characterization of DESHIMA have recently been reported\cite{E1}. Here we will focus on the on-telescope measurements.

\begin{figure*}[htbp]
 \includegraphics[width=\textwidth]{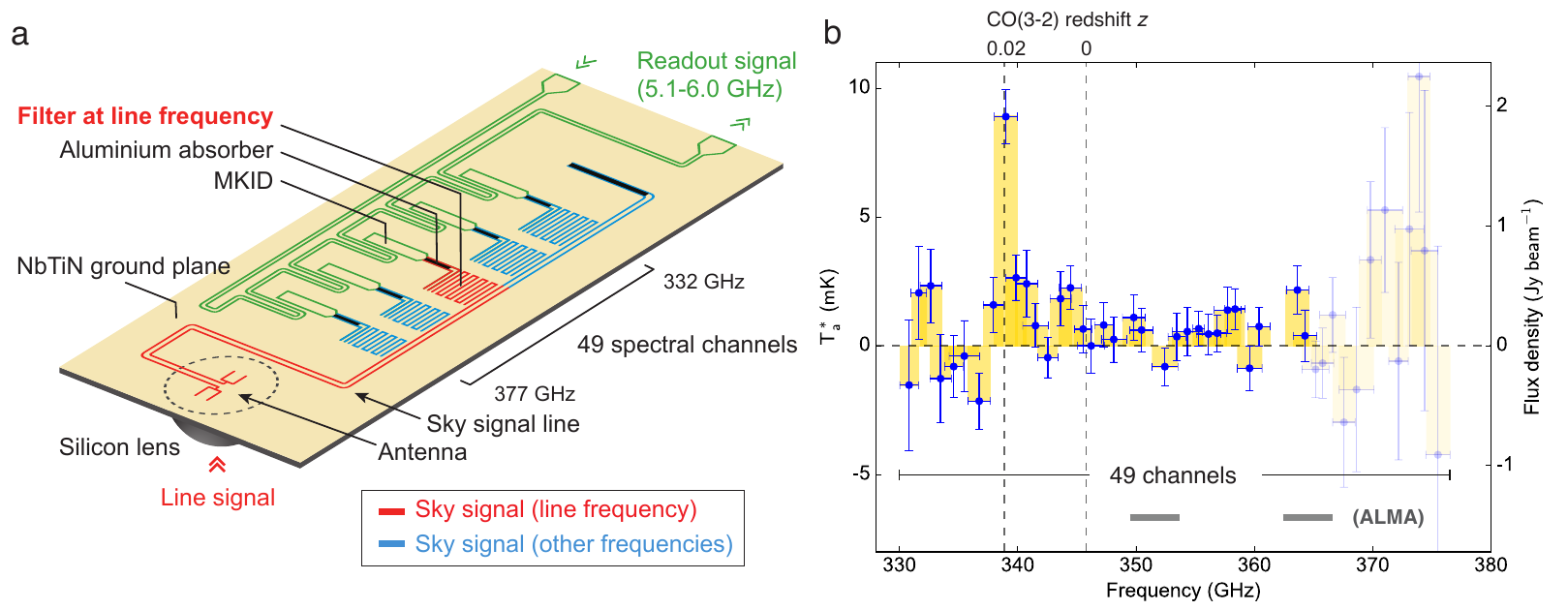}
 \caption{\label{fig:concept}
 \linespread{1.0}\selectfont
\textbf{Integrated superconducting spectrometer detection of redshifted CO(3--2) line emission from the luminous infrared galaxy VV~114.}
\textbf{a}, Schematic of the integrated superconducting filterbank (ISS) chip of DESHIMA. The flow of a redshifted CO signal (339 GHz) is indicated in red, whereas the flow of signal at all other frequencies is indicated in blue. The filterbank consists of 49 spectral channels, of which five are depicted here. Each spectral channel consists of a bandpass filter and an MKID. Only the filter with a passband that matches the redshifted astronomical line frequency resonates, and delivers power to the MKID at its output. The MKID measures the amount of signal power that is absorbed in the aluminium section (black). The astronomical signal line is terminated with a coplanar waveguide (CPW) with an aluminium center strip (black) to prevent reflective standing waves by absorbing the remaining power. The flow of the readout signal (5.1--6.0~GHz) is indicated in green. All MKIDs are read out simultaneously through a common microwave readout CPW.
\textbf{b}, Spectrum of VV~114 measured with the ISS. The response of each channel of the ISS is plotted as a function of the peak frequency of the response curves presented in Fig.~\ref{fig:deshima}i. The previously reported peak frequency of CO is indicated\cite{2008ApJS..178..189W}. 
The horizontal error bars and the yellow shades under them indicate the full width at half maximum of the channel response. The vertical error bars indicate the 1$\sigma$ noise level per channel. 
The right-side vertical axis is calculated from antenna temperature $T_\mathrm{a}^*$ assuming frequency-independent values of $\eta_\mathrm{MB}=0.34$  for the main beam efficiency and $\Omega_\mathrm{MB}=1.9\times 10^{-8}$ str for the main beam solid angle. The channels above 365 GHz have a low signal-to-noise ratio because of the low atmospheric transmission. The instantaneous bandwidth of ALMA band 7 is indicated for comparison\cite{6111333}.
}
\end{figure*}

\begin{figure*}
\includegraphics[width=0.8\textwidth]{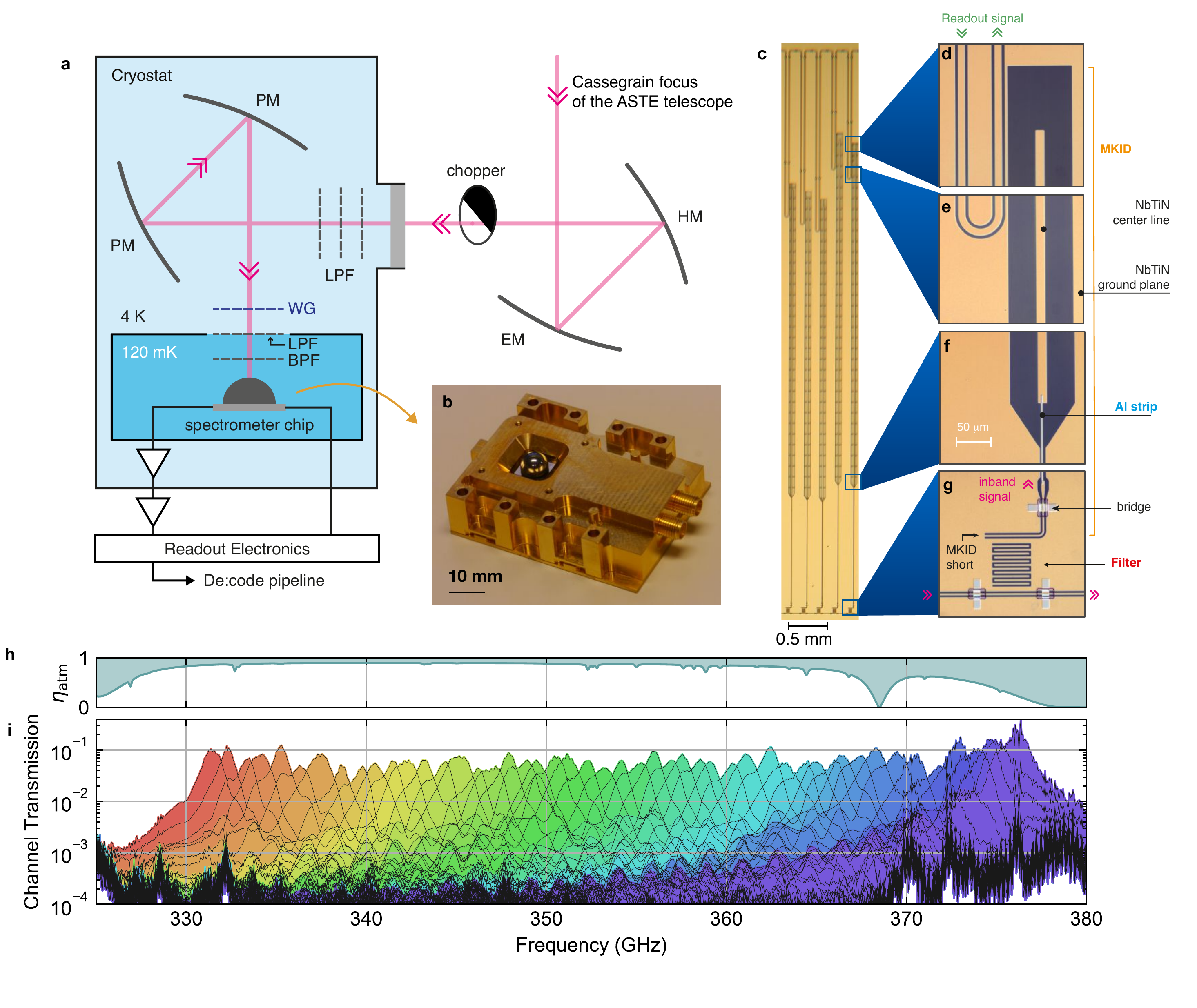}
\caption{\label{fig:deshima} 
\linespread{1.0}\selectfont
\textbf{DESHIMA spectrometer system in the ASTE telescope cabin.}  
\textbf{a}, The submm signal from the ASTE telescope is coupled to the DESHIMA cryostat window with an ellipsoidal tertiary mirror (EM) and a hyperbolic quaternary mirror (HM) in a cross-Dragonian configuration. Inside the cryostat, the signal is guided to the ISS chip through an optics tube, which consists of two parabolic mirrors and a stack of low-pass filters (LPF) and baffles, a wire grid (WG) polarizer, and a final bandpass filter (BPF). The response of all MKIDs are read out simultaneously at 160~Hz through one pair of coax lines that are connected to the RF readout electronics\cite{E1}. 
\textbf{b}, The ISS chip in a copper housing. The Si lens is visible.  
\textbf{c}, Micrograph of a subset of five channels of the filterbank.
\textbf{d-g}, Further close-up view of one of the spectral channels of the filterbank. The filter (g) is an interdigitated pattern etched in NbTiN, acting as a resonant bandpass filter. On both sides of the filter, the ground planes of the sky signal line are connected with aluminium bridges over a block of UV-patterned polyimide. Every MKID (d-g) consists of an end-shorted NbTiN CPW that couples to the filter (g), a narrow CPW with an aluminium center line (f, g), and an open-ended wide NbTiN CPW (d-f). The MKID is coupled to the microwave readout line near the open end (d, e).
\textbf{h}, Atmospheric transmission at zenith as a function of frequency, calculated\cite{2001ITAP...49.1683P} assuming a precipitable water vapor of 0.5 mm.
\textbf{i}, Spectral response of all 49 filter channels of the filterbank,  normalized to the signal power at the entrance of the filterbank. The response (vertical axis) was measured while sweeping the frequency (horizontal axis) of a narrow-band signal which was radiated into the spectrometer from a photomixing submm-wave source\cite{E1}.
}
\end{figure*}

The design and working principle of the DESHIMA ISS chip is illustrated in Fig.~\ref{fig:concept}, using the first detection of a redshifted extragalactic emission line using this technique from VV~114, a luminous infrared galaxy (LIRG). First, the lens-antenna on the chip receives the astronomical signal from the telescope and optics, and couples it to a small transmission line. The signal then enters the filterbank section that consists of 49 spectroscopic channels. Micrographs of a few spectral channels are presented in Fig.~\ref{fig:deshima}c-g.  Each filter is a superconducting submm-wave resonator, with a resonance frequency that sets the peak of its passband. Because the signal from VV~114 contains only a single strong CO(3--2) line, only one filter intercepts the signal and delivers power to the MKID at its output. The responding channel has a passband of 1.0 GHz around 339.0 GHz, which is consistent with the CO(3--2) rest frequency of 345.796 GHz and the redshift $z=0.02$ of VV~114\cite{2008ApJS..178..189W}. 

We evaluated DESHIMA on the ASTE telescope in the period from October to November 2017.  The layout of DESHIMA in the ASTE cabin is schematically presented in Fig.~\ref{fig:deshima}a. Before the measurements on sky, we verified that the instrument optical sensitivity of DESHIMA in NEP is not affected by the ASTE cabin environment, using the same hot-cold measurement technique as used in the laboratory tests\cite{E1}. The response of the MKIDs to the sky signal was calibrated and linearized using skydip measurements (see Methods: `Calibration of the sky signal response'). The telescope beam shape and main beam efficiency were measured on Mars (see Methods: `Beam efficiency'). 

We demonstrate the key applications of this instrument, as a redshift machine and as a fast multi-line spectral imager of large areas, by utilizing the on-sky measurements, which we also analyze to demonstrate the sensitivity achieved. The first measurement of a cosmologically redshifted molecular line with an instrument of this type is shown in Fig.~\ref{fig:concept}b. The width of the line is comparable ($\sim$0.5~GHz\cite{2008ApJS..178..189W}) to the spectrometer resolution, which is an optimum condition for achieving both high sensitivity and a wide instantaneous band for a fixed number of detectors (see Methods: `Optimum frequency resolution'). Using a combination of a chopper wheel and slow (0.5~Hz) nodding of the ASTE telescope (see Methods: `Position switching observations'), we obtained a CO line signal-to-noise ratio ($\mathrm{SNR}$) of $\sim$9 in an on-source integration time of 12.8 min. This method can be applied to targeted, wideband multi-line spectroscopy of high-$z$ SMGs to identify their redshift and study their emission line spectra.

\begin{figure*}[htbp]
 \includegraphics[width=0.8\textwidth]{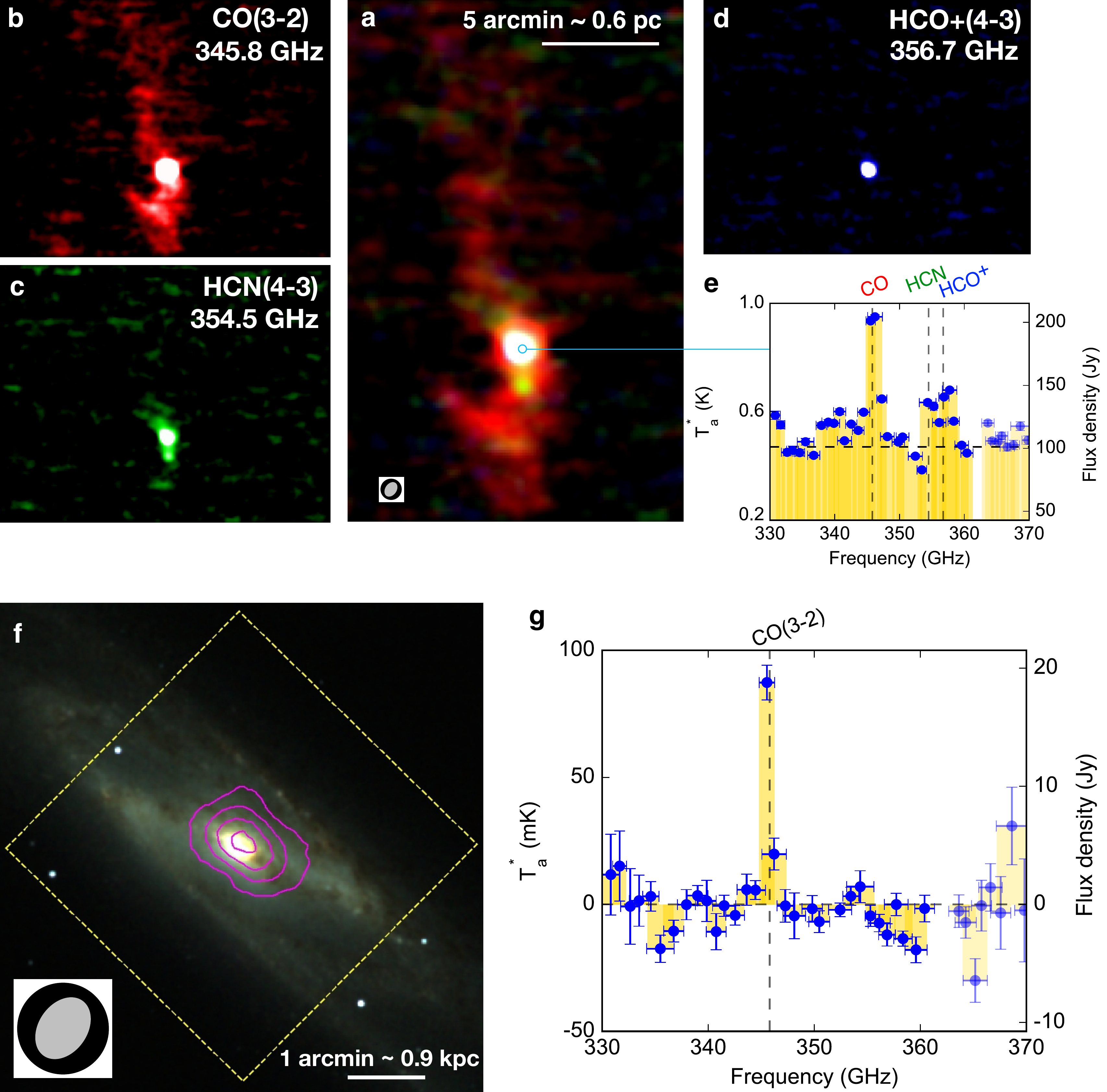}
 \caption{\label{fig:orion} 
 \linespread{1.0}\selectfont
 \textbf{DESHIMA spectral maps of the Orion nebula and the barred-spiral galaxy NGC~253.}  
 \textbf{a}, DESHIMA CO(3--2)/HCN(4--3)/HCO$^+$(4--3) RGB image of the Orion nebula. The box at the bottom left shows the size of the 43$''$-diameter circular aperture used for photometry (black), together with the best-fit two-dimensional Gaussian beam at half-power level (gray). The ellipse is set at the typical position angle of the beam, which rotated by up to $\sim\pm$45$^\circ$ with respect to the map during the observations of both Orion and NGC~253 (panel f). The effective resolution of the images shown in panels a-d and f is  $\sim$39$''$, given by the convolution of the aperture and the beam. 
 \textbf{b,c,d}, Individual DESHIMA CO(3--2), HCN(4--3) and HCO$^+$(4--3) maps used in the RGB map. The images are continuum-subtracted. 
 \textbf{e}, DESHIMA spectrum of the Orion KL region based on aperture photometry. The spectrum is plotted in the same manner as the spectrum in Fig \ref{fig:concept}b. (Many of the vertical error bars are smaller than the points.) The horizontal dashed line indicates the continuum flux at each position derived by averaging assumed, emission-free channels.
\textbf{f}, DESHIMA CO(3--2) map (contours) of NGC~253 on a 2MASS $JHK$ RGB image. The contour levels are 3, 6, 9, and 12$\sigma$. The position offset between the CO-brightest point and the brightest point in the 2MASS image is $\sim$4.5$''$, which is comparable to the typical pointing error of the observations presented in this Letter. The yellow dashed line shows the coverage of the DESHIMA observations. \textbf{g}, The DESHIMA spectrum for the CO-brightest point of NGC~253 based on aperture photometry. The 1$\sigma$ noise level is derived from the map, after smoothing with the 43$''$-diameter circular aperture.
}
\end{figure*}

The second result is the successful acquisition of wideband spectral maps using on-the-fly (OTF) mapping. In Fig.~\ref{fig:orion}a we show a three-color composite map of the Orion nebula, which combines channel maps of CO(3--2), HCN(4--3), and HCO$^+$(4--3), as presented in Fig.~\ref{fig:orion}b-d. The OTF map captures the extended structure of the CO line\cite{2016MNRAS.457.2139C}, whereas the HCN and HCO$^+$ lines are more localized. In making the line intensity maps, the signal component common to all channels was subtracted as the baseline `continuum', as indicated by the horizontal dashed line in the spectrum presented in Fig.~\ref{fig:orion}e. Because this component contains signal from many emission lines unresolved by DESHIMA, we complement this result with a spectral map of the nearby barred-spiral galaxy NGC~253, which exhibits CO(3--2) as a single dominant emission line with clear line-free channels around it. The map of NGC~253 also has well defined emission-free positions in the direction vertical to its disk. The DESHIMA CO(3--2) map of NGC~253 captures the extended emission along the bar\cite{2001A&A...373..853D}. The Orion and NGC~253 maps together show that ISSs can be operated in OTF mode, by removing fluctuations of the atmosphere and the instrument in a manner similar to observations with coherent spectrometers. 

\begin{figure*}
 \includegraphics[width=1\textwidth]{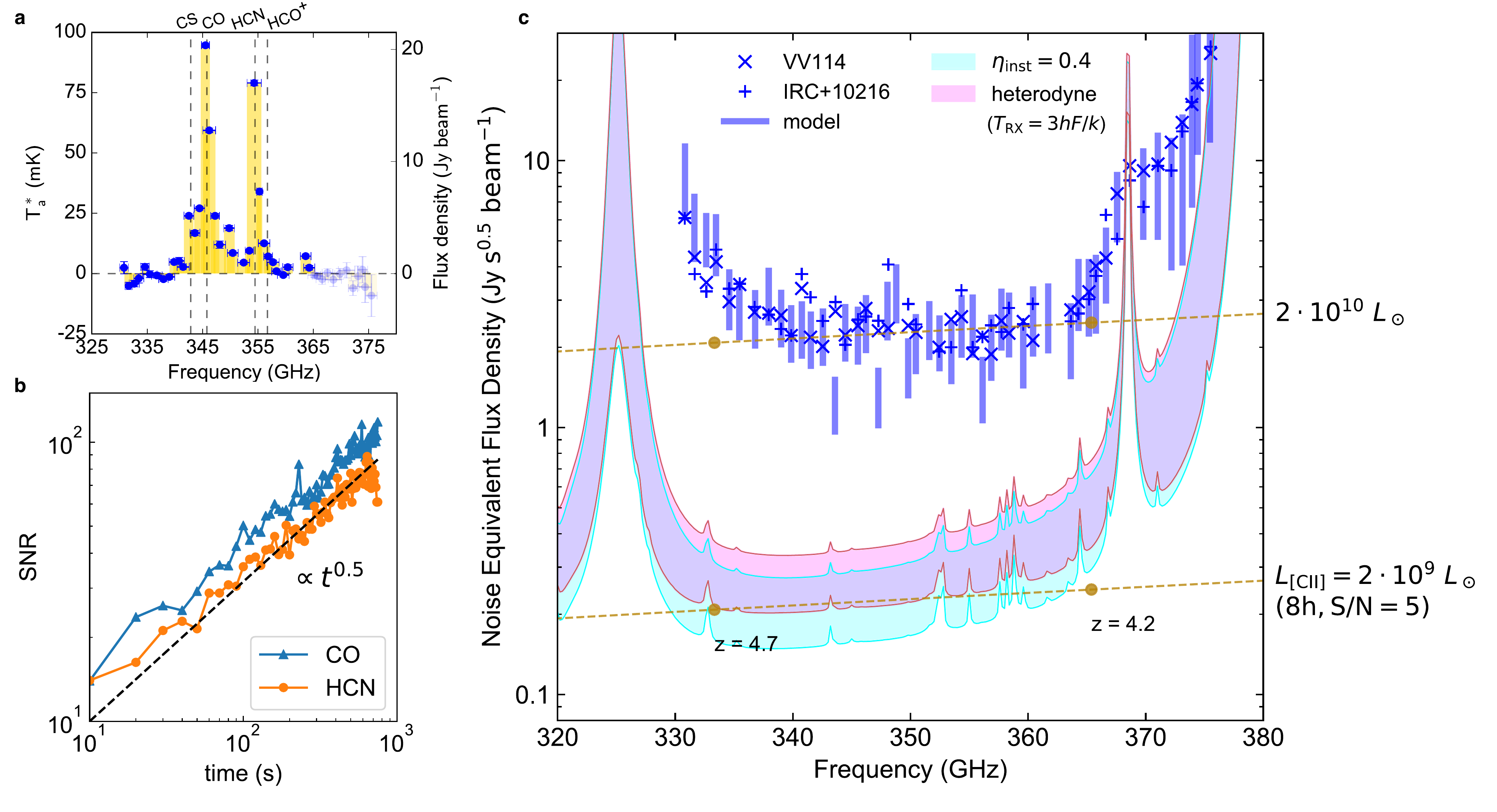}
 \caption{\label{fig:NEFD} 
 \linespread{1.0}\selectfont
 \textbf{Foreground photon-noise limited sensitivity of the ISS and its fundamental limits.} 
 \textbf{a}, Broadband DESHIMA spectrum of IRC+10216. The spectrum is plotted in the same manner as the spectrum in Fig \ref{fig:concept}b. (Many of the vertical error bars are smaller than the points.)
\textbf{b}, Signal to noise ratios of the HCN(4--3) and CO(3--2) lines as functions of on-source integration time. The dashed line is a guide to the eye, showing a slope of $t^{0.5}$. 
 \textbf{c}, Noise equivalent flux density (NEFD) of DESHIMA. The symbols `$\times$' and `$+$' show the NEFD of each frequency channel of DESHIMA measured during the observation of VV~114 and IRC+10216, respectively. The blue solid bars are the theoretical NEFD for the photon-noise limited case (see Methods: `Sensitivity'). The span in the model NEFD indicates the range of NEFD for a precipitable water vapor (PWV) in the atmosphere in the range of 0.5--1.5 mm. The area shaded in cyan represents a possible future improvement in which the instrument optical efficiency is improved from 0.02 (this work) to 0.4, and the telescope aperture efficiency is improved from 0.17 (see Methods: `Beam efficiency') to 0.4. For comparison, the area shaded in pink indicates the sensitivity of a coherent receiver that has a receiver noise temperature of $T_\mathrm{RX}=3hF/k\sim 50\ \mathrm{K}$. We have assumed the same spectral resolution in the calculation of the shaded areas, corresponding to a filter $Q=F/\mathrm{FWHM}$ of 300 that is equal to the average $Q$ factor of the filters of DESHIMA\cite{E1}. The dashed curves indicate the NEFD level required to detect redshifted [C\,\textsc{\smaller II}] lines with line luminosities as indicated, at a signal to noise ratio of 5 in 8 hours.
}
\end{figure*}

The sensitivity of DESHIMA has been measured from the observation of the post-asymptotic giant branch (AGB) star IRC+10216. This source exhibits a strong HCN(4--3) line that is spatially unresolved with the DESHIMA/ASTE beam. After an on-source integration time of $ t_\mathrm{on}\sim10^3$~s, a HCN line SNR of $\sim$67 was reached, as presented in Fig~\ref{fig:NEFD}a,b. The $\mathrm{SNR}\propto t_\mathrm{on}^{0.5}$ dependence shows good stability during integration. The noise equivalent flux density (NEFD) per channel has been estimated from this data set, and is presented in Fig.~\ref{fig:NEFD}c. For the frequency range in which the atmosphere is most transparent (see Fig.~\ref{fig:deshima}h), a NEFD of $\sim$2--3~$\mathrm{Jy\ s^{0.5}\ beam^{-1}}$ is reached. The NEFD inferred from the observation of VV~114 is similar, as can be seen in Fig.~\ref{fig:NEFD}c, confirming that the estimation depends little on the observing conditions or the properties of the source. This sensitivity would allow for example a 5$\sigma$ detection of a [C\,\textsc{\smaller II}] line from a hyper-luminous infrared galaxy (HyLIRG) at redshift 4.2--4.7, with an on-source integration time of 8 hours, as indicated in the figure. Furthermore, the blue bars in Fig.~\ref{fig:NEFD}c indicate the on-sky NEFD predicted from the optical efficiency of DESHIMA measured in the laboratory\cite{E1}, in combination with the aperture efficiency we measured on Mars in this work (see Methods: `Beam efficiency'). 

The excellent agreement between the instrument design, laboratory sensitivity, and on-sky sensitivity shows that DESHIMA on ASTE reaches the foreground photon-noise limit. This means that the sensitivity is limited only by the foreground photon noise and by the coupling efficiency between source and detector. 
 The limiting factors here are the ISS chip design and the intrinsic coupling between the warm optics and the chip. 
 The efficiency of the chip is currently $\sim$0.08, due to the design of the coplanar filters and the oversampling\cite{E1}.
 This can be improved to $\sim$0.5 by adopting microstrip filters\cite{Wheeler:2016dr} based upon amorphous silicon: We recently measured a loss tangent of tan$\delta = 10^{-4}$ at $\sim$350 GHz (S. H\"ahnle, private communication). 
 Additionally a more advanced filter geometry is needed to couple more than 50\% of power into a single filter: an example would be to couple the power from several over-sampled filters\cite{Shirokoff:2012fx} into a single MKID\cite{Bueno2018}. 
 These developments\cite{Wheeler:2016dr,Shirokoff:2012fx,Bueno2018} provide a path to improving the chip efficiency. Regarding the optics, a careful selection of the quasioptical filters, and using isotropic substrates (e.g., silicon), can bring the transmission from the window to the on-chip antenna feed point from the current 0.22 to $\sim$0.5. As indicated in Fig.~\ref{fig:NEFD}c, an instrument optical efficiency of 0.4 and a telescope aperture efficiency of 0.4 would allow easy detection of the unlensed ultra-luminous infrared galaxy (ULIRG) population at $z=4.2\text{--}4.5$ with the [C\,\textsc{\smaller II}] line. 
In this case the full system sensitivity on sky becomes comparable to a state-of-the-art heterodyne receiver instrument\cite{Ito:2018bb}, because both systems are limited mainly by the atmosphere. 
The ISS technology is highly scalable towards ultra-wide bandwidths and many spatial pixels. Half-wave microstrip resonators\cite{Shirokoff:2012fx} are intrinsically capable to be used as filters in a 1:2 bandwidth spectrometer coupled to a wideband antenna\cite{OBrient:2013hc,2017ApPhL.110w3503B}, a similar filter design with an open and a shorted end could be used in a 1:3 bandwidth. With the current density of channels in the filterbank, a 500 channel filterbank covering a 1:3 instantaneous bandwidth (e.g., 240-720~GHz) at a resolution of $F/\Delta F=500$ would still be as small as $\sim$5~$\mathrm{cm^2}$. The wide instantaneous bandwidth and sensitivity will easily allow simultaneous detection of multiple emission lines (e.g., CO, [C\,\textsc{\smaller II}])\cite{2011ITTST...1..241S} that is required to determine an unambiguous spectroscopic redshift. Furthermore, 3D integral field spectrographs\cite{2018Natur.562..229W} can be formed naturally with a 2D array of such spectroscopic pixels\cite{2016SPIE.9906E..26K}. Such an instrument will allow the exploration of cosmic large-scale structures with an unprecedented sensitivity and spatial scales, depicting the 3D distribution of galaxies with abundant molecular and atomic lines across the cosmological volume and time\cite{2019arXiv190304779G,2017arXiv170902389F,2016SPIE.9906E..26K,2018SPIE10700E..5XP}.
 
 \clearpage
 
\clearpage 

\section*{Methods}

\subsection*{Calibration of the sky signal response}

Conversion from the relative frequency response of the MKID to the line-of-sight brightness temperature of the sky ($T_\mathrm{sky}$) is based on a model that uses the atmospheric transmission measured by DESHIMA itself. 
We conducted fast and wide scans of the telescope elevation (`skydip' observations) 22 times throughout the observing session, with an elevation range of 32$^\circ$--88$^\circ$. The PWV values were typically in the range of 0.4--3.0 mm, with a mean value of 0.9 mm, according to the water vapor radiometers mounted on each telescope of the Atacama Large Millimeter/submillimeter Array (ALMA)\citep{2013A&A...552A.104N}, located in the vicinity of ASTE.  We define $x$ as the fractional change  in MKID readout resonance frequency $f$, from when the instrument window is facing the blackened absorber on the chopper wheel at ambient temperature $T_\mathrm{amb}$ (Fig. \ref{fig:deshima}a), to when the instrument looks at the sky with a brightness temperature $T_\mathrm{sky}$ thorough the telescope optics: $ x \equiv \{f(T_\mathrm{sky})-f(T_\mathrm{amb} )\}/f(T_\mathrm{amb})$. $T_\mathrm{sky}$ was calculated from
\begin{equation}
 T_\mathrm{sky} = \eta_\mathrm{fwd} (1-\eta_\mathrm{atm}) T_\mathrm{atm} + (1-\eta_\mathrm{fwd}) T_\mathrm{amb},
\end{equation}
where $\eta_\mathrm{fwd}=0.93$ is the telescope forward efficiency of ASTE. $\eta_\mathrm{atm}$ is the transmission coefficient of the atmosphere calculated\citep{2001ITAP...49.1683P} from the PWV and telescope elevation, taking into account the frequency response of each channel (Fig. \ref{fig:deshima}i). We have assumed the physical temperature of the atmosphere $T_\mathrm{atm}$ to be equal to the outside ambient temperature $T_\mathrm{amb}$ measured with the weather monitor at the ASTE site. From a least-$\chi^2$ fitting to the square-law $T_\mathrm{sky} \propto x^2$ dependence for an aluminium-based MKID response\cite{E1}, we obtain a calibration model curve as shown in Supplementary Fig. S1b. We perform an iterative estimation of the PWV using all channels of the filterbank with $\eta_\mathrm{atm}$, and from that the PWV, as a single common free parameter. In this way we obtain a calibration model that has a dispersion of $\sim$4\% for all skydip measurements. The error is significantly smaller than the initial model that directly uses the PWV from the ALMA radiometer data (Supplementary Fig. S1a). This  calibration model is used for all astronomical measurements reported in this Letter.

\subsection*{Beam efficiency}

We used Mars with an apparent diameter of 3.99$''$ and a brightness temperature $T_\mathrm{brightness} = 210$ K\cite{Butler2012} to measure the beam pattern and efficiency of the DESHIMA optics coupled with the ASTE 10 m telescope. The data were obtained at 12:48 UTC, 2017 November 15 (daytime in Chile) with a  precipitable water vapor (PWV) of 1.8 mm. The intensity was calibrated to antenna temperature $T_\mathrm{a}^\ast$ using a standard chopper wheel method\cite{ToolsOfRadioAstronomy}. The flux-scaling, noise removal and map-making were performed using a data analysis software  De:code (DESHIMA Code for data analysis)\cite{decode}. Considering the accuracy in responsivity calibration ($\sim$4\%), chopper wheel calibration together with the uncertainty in $\eta_\mathrm{MB}$ (10--15\%), and the accuracy of the planet model ($\sim$5\%), the absolute flux accuracy is estimated to be 12--17\%. The main beam shape is measured by fitting a 2-dimensional Gaussian to the 350 GHz continuum image on Mars (Supplementary Fig. S2). The source-deconvolved beam size is estimated to be 31.4$''$ $\pm$ 2.8$''$ by 22.8$''$ $\pm$ 3.1$''$ (in full width at half maximum, FWHM) with a position angle of 145.4$^\circ$. We estimate the main-beam efficiency by comparing the peak intensity with what is expected from the model and find $\eta_\mathrm{MB}$ = 0.34 $\pm$ 0.03 at 350 GHz (see Supplementary Note 1 for details).
The main beam solid angle $\Omega_\mathrm{MB}$ and main beam efficiency yield an aperture efficiency of $\eta_\mathrm{A}=(\lambda^2/A_\mathrm{p})\cdot(\eta_\mathrm{MB}/\Omega_\mathrm{MB})=0.17$, 
where $A_\mathrm{p}$ is the physical area of the ASTE primary mirror and $\lambda$ is the wavelength. 
This value is much lower than previous 350 GHz measurements with a heterodyne receiver on ASTE ($\eta_\mathrm{MB} \sim 0.6$)\cite{Ito:2018bb}, and can be attributed to an offset of the instrument beam of DESHIMA. In a post analysis we have taken the beam pattern from the cryostat, measured in phase and amplitude\cite{E1}, and propagated it using Zemax\cite{Zemax} to estimate the illumination pattern on the surface of the ASTE mirrors. Supplementary Fig. S2c shows the resulting far-field beam pattern, which explains both the oval beam shape and the aperture efficiency of 0.17.

\subsection*{Position switching observations}

For the single-pointing observations of IRC+10216 and VV~114, 
we oscillated the pointing of the ASTE telescope between 
the source position and a position 60$''$ away from the source position in the azimuth direction, with a 2~s duty cycle. 
We integrated the spectrum of the target source that is contained within a circle of 11$''$ radius (on-source position), which corresponds to an approximate half width at half maximum of the ASTE beam.
We regarded the data beyond 27$''$ from the target as off-source positions.
The data were continuously recorded during the scans with a sampling rate of 160~Hz.\cite{Rantwijk2016}
Because the frequency of the telescope nodding is lower than the typical onset of $1/f$ noise of the detectors\cite{E1} ($\sim$1~Hz, corresponding to an Allan variance time of $\sim$1 s), the time-stream data are calibrated at 10~Hz using a blackened absorber on the rotating chopper wheel placed in front of the receiver (Fig. \ref{fig:deshima}a).
We took the difference 
$T_\mathrm{sky}-T_\mathrm{amb}$ at on- and off-source positions and used the standard chopper wheel calibration method\cite{ToolsOfRadioAstronomy} to correct for atmospheric absorption, and to convert $T_\mathrm{sky}$ to antenna temperature $T_\mathrm{a}^\ast$. Throughout the paper, on-source integration time refers to the total time that DESHIMA was observing the on-source position, excluding overheads for calibration.

\subsubsection*{IRC+10216}

The broadband spectra of the post-AGB star IRC+10216 were taken on 16--20 November 2017.
The PWV measured by ALMA was typically 0.75~mm.
The observed data were reduced using De:code\cite{decode}.
After the chopper wheel calibration as mentioned above, the strong continuum emission of the target was removed by subtracting the median baseline, which was estimated in the frequency range of $<$340~GHz and $>$360~GHz.

The obtained broadband spectrum is shown in Fig.~\ref{fig:NEFD}a.
The noise level of each spectral channel is determined by applying an iterative integration method with random sign inversion (the jack-knife method, hereafter. See Supplementary Note 2 for details). 
Two remarkable peaks are found at $\sim$345~GHz and $\sim$354~GHz, corresponding to CO(3--2) and HCN(4--3) lines.
The peak intensities are 94.7~mK and 79.0~mK in $T_\mathrm{a}^\ast$.
The integrated intensities of the CO and HCN lines are estimated by integrating over 340--350~GHz and 353--358~GHz, to be 247 and 104~mK~km~s$^{-1}$, respectively. 
The spectral shape agrees with that expected from a spectral survey observation with the Submillimeter Array\cite{2011ApJS..193...17P}.
The integrated intensities of the CO and HCN lines are measured to be 58\% and 31\% of those measured by a previous observation using the Caltech Submillimeter Observatory (CSO) 10~m submillimeter telescope, after correcting for the spectral resolution, the beams size of DESHIMA on ASTE measured on Mars, and the main beam efficiency of $\eta_\mathrm{MB}=0.34$. These intensity ratios indicate that the main beam efficiency of DESHIMA/ASTE should be $\eta_\mathrm{MB}\sim$0.20--0.38, which is consistent with our measurement using Mars, and the Zemax simulation as explained in the `beam efficiency' section. 

A plot of the signal-to-noise ratio (SNR) of the CO and HCN lines as functions of on-source integration time is presented in Fig. \ref{fig:NEFD}b. The SNR was calculated by dividing the signal by the noise level of that single spectral channel (corresponding to the vertical error bars in the spectrum as shown in Fig.~\ref{fig:NEFD}a but for different integration times.)  For calculating the NEFD in Fig. \ref{fig:NEFD}c, we have adopted the HCN line over the CO line, because the HCN line is more concentrated near the target\cite{1992A&A...266..365W, 1994ApJS...95..503W}. 
The noise level of the HCN channel is estimated to be 1.2~mK after an on-source integration time of 12.6~min, corresponding to a NEFD of  3.2~Jy~beam$^{-1}$~s$^{0.5}$.

\subsubsection*{VV~114}

The interacting galaxy pair VV~114 has been observed on 16 and 21 November 2017.
The typical PWV measured by ALMA were 0.7~mm on the 16th and 0.9~mm on the 21st.
The scan pattern and data reduction method were the same as for IRC+10216;
the continuum emission was removed by subtracting a median baseline of the spectrum, estimated in the frequency range of $<$335~GHz and $>$345~GHz.

The broadband spectrum of VV~114 is shown in Fig.~\ref{fig:concept}b.
A significant emission line is detected at 339~GHz, corresponding to the redshifted CO(3--2) spectrum\cite{2008ApJS..178..189W}.
The peak intensity of the CO line is 8.9~mK in $T_\mathrm{a}^\ast$ and the integrated intensity is estimated by integrating over 337.9--341.5~GHz to be 12.4~K~km~s$^{-1}$.
Adopting the beam size of DESHIMA on ASTE measured on Mars and the main beam efficiency of $\eta_\mathrm{MB}=0.34$, the total flux density of the CO(3--2) emission is estimated to be 2.44$\times10^3$~Jy~km~s$^{-1}$.
Regarding the difference in the integrated regions on sky, the estimated flux density is in reasonable agreement with the previous estimate with the JCMT (2956$\pm$133~Jy~km~s$^{-1}$)\cite{2008ApJS..178..189W}.
No other line feature is found except for tentative detections near 332~GHz and 343~GHz.
The frequencies of these features are consistent with the redshifted frequencies of methanol in transitions of 7(1,6)--6(1,5)E and 4(0,4)--3(-1,3)E, respectively.
The noise level of the spectrum is typically 1.0~mK in $T_\mathrm{a}^\ast$ with an integration time of 12.6~min, which is equivalent to a NEFD of 2.5~Jy~beam$^{-1}$~s$^{0.5}$.
The deep integration and the simple shape of the spectrum allow us to estimate the NEFD at each channel with the jack-knife method as shown in Fig. \ref{fig:concept}b.
We find that the NEFDs of all channels except for some higher frequencies achieve a good agreement with that of the theoretical prediction, as presented in Fig. \ref{fig:NEFD}c.

\subsection*{On-the-fly mapping observations}

On-the-Fly (OTF) mapping observations toward the Orion KL region and NGC~253 (Fig.~\ref{fig:orion}) were performed with spatial raster scans, in which the DESHIMA/ASTE beam ran linearly across the area of interest at a constant speed. The data taken at each side of the scans were used to subtract the foreground sky emission.
Absolute flux calibration was performed with the standard chopper wheel method\cite{ToolsOfRadioAstronomy} at the beginning of each mapping observation. 
The signal spectra and the noise on each channel map were obtained by aperture photometry on the map, adopting an aperture diameter of 43$''$.

\subsubsection*{Orion}
DESHIMA observations towards the massive star-forming region around Orion KL were executed on 8--12 November 2017. The area presented in Fig.~\ref{fig:orion}b--d was divided into six sub-regions, which each have a size of 29$'$ by 4$'$. After 28 separate observations of these sub-regions in a total on-source time of 12.5 hours, the data were combined to produce the final map of 29$'$ by 22$'$. After the basic data reduction process described above, 
we model the common signal across all channels as a superposition of continuum emission from the source and sky foreground emission, and remove the sky contribution based on an inter-channel correlation analysis.
 We applied a moderate high-pass filter in the image domain in the scanning direction to remove part of the instrument and atmospheric $1/f$ noise, because we did not continuously rotate the chopper during this observation.
Finally, we convolved the map with a 43$''$-diameter aperture to obtain the maps of CO, HCN, and HCO$^+$ presented in Fig.~\ref{fig:orion}b--d. The spectrum at the point of Orion KL is displayed in Fig.~\ref{fig:orion}e.

\subsubsection*{NGC~253}
The OTF spectral map of the nearby barred-spiral galaxy NGC~253, presented in Fig.~\ref{fig:orion}f,g, was taken on 6--7 November 2017. The map size is 4.2$'$ by 4.2$'$. The total on-source time was 1.7~hours. The data were reduced in the same manner as the Orion KL data, except that we used the median value of all channels to subtract the foreground sky emission. No continuum emission was detected from NGC~253, in an analysis similar to that for Orion. The obtained 1$\sigma$ noise is typically $\sim$5.3~mK from 335~GHz to 365~GHz. The map of the 345.53~GHz channel shows a 12.8$\sigma$ detection of CO(3--2) emission from NGC~253, as shown in Fig.~\ref{fig:orion}f where the DESHIMA CO(3--2) map is overlaid on a 2MASS $JHK$ RGB image\cite{2006AJ....131.1163S}. The spectrum at this point, for a 43$''$ aperture, is shown in Fig.~\ref{fig:orion}g. 
The total integrated intensity of the CO(3--2) emission at the peak position for a 26.7$''$ aperture is estimated to be 138~K~km~s$^{-1}$.
This value is consistent with the previous measurement using a heterodyne receiver installed on the CSO 10~m telescope\cite{2004A&A...427...45B}, taking into account the difference in beam size, and the accuracy of the absolute flux calibration. The total integrated intensity of the CO(3--2) spectrum taken with CSO is estimated to be $815.6\pm 163.1$~K~km~$\mathrm{s}^{-1}$\cite{2004A&A...427...45B}. Since the beam size of CSO is 21.9$''$, which is smaller than the DESHIMA beam, it should be corrected for comparison. If we assume that the emission is uniformly distributed over the DESHIMA beam, the total integrated intensity is corrected to be 205~K~km~$\mathrm{s}^{-1}$ for a 21.9$''$ aperture. Adopting a main beam efficiency of 0.34, the total integrated intensity is thus 643~K~km~$\mathrm{s}^{-1}$. This is slightly lower than the CSO value but within the margin of the uncertainty in absolute flux calibration and the distribution of source emission.

\subsection*{Sensitivity}

The power coupled to the MKID is the sum of the power from the sky and the power from warm spillover. This is given by
\begin{equation}
\begin{split}
P_\mathrm{MKID} &= \{\eta_\mathrm{fwd}(1-\eta_\mathrm{atm})+(1-\eta_\mathrm{fwd})\}\eta_\mathrm{inst}B(F,T_\mathrm{amb})\Delta F,\\
\end{split}
\end{equation}
where $\eta_\mathrm{fwd}=0.93$ is the forward efficiency of the telescope, i.e. all power coupled to the sky, $\eta_\mathrm{inst}$ is the coupling from the cryostat window to the MKID detector, and $\eta_\mathrm{atm}$ is the atmosphere transmission given by the PWV and elevation. In Fig. \ref{fig:NEFD}c, we have adopted $\eta_\mathrm{inst}=0.02$ for the model bars, based on the laboratory test\cite{E1}. $B(F, T_\mathrm{amb})$ is the single polarization Planck brightness at frequency $F$ and temperature $T_\mathrm{amb}$, taken identical for the ground, cabin and atmosphere. The photon noise limited noise equivalent power ($\mathrm{NEP_\mathrm{ph}}$), at the detector, is given by\cite{E1}
\begin{equation}
\label{eq:NEPph}
\begin{split}
\mathrm{NEP_{ph}} &= \sqrt{2P_\mathrm{MKID}(hF+P_\mathrm{MKID}/\Delta F)+4\Delta_\mathrm{Al} P_\mathrm{MKID}/\eta_\mathrm{pb}}.
\end{split}
\end{equation}
Here, $h$ is the Planck constant, $\Delta F$ is the effective bandwidth of the filter channel, $\Delta_\mathrm{Al}=188\ \mathrm{\mu eV}$ is the superconducting gap energy of aluminium, and $\eta_\mathrm{pb}\sim 0.4$ is the pair-breaking efficiency\cite{2014SuScT..27e5012G}. 
The foreground photon-noise limited noise equivalent flux density ($\mathrm{NEFD_{ph}}$), evaluated on-sky, has to take into account the instrument coupling, aperture efficiency $\eta_\mathrm{A}$, and the physical area of the ASTE telescope $A_\mathrm{p}$, and is given by
\begin{equation}
\begin{split}
\mathrm{NEFD_{ph}} &= \frac{\mathrm{NEP_{ph}}}{\sqrt{2}\eta_\mathrm{pol}\eta_\mathrm{inst}\eta_\mathrm{fwd}\eta_\mathrm{atm}\eta_\mathrm{A}A_\mathrm{p}\Delta F}.
\end{split}
\end{equation}
Here, the factor $\sqrt{2}$ accounts for the NEP being defined for 0.5 s integration time, and
$\eta_\mathrm{pol}=0.5$ accounts for the fact that DESHIMA is sensitive to a single linear polarization.

\subsection*{Optimum frequency resolution}

Here we will show that the signal-to-noise-ratio (SNR) for a single ISS channel matched to the center frequency of an astronomical line is maximum when the channel frequency width $\Delta F_\mathrm{ch}$ is equal to the line width $\Delta F_\mathrm{line}$, under the assumption that the measurement is limited by the foreground/background photon-noise. In the following analysis we adopt a frequency width ratio $r \equiv \Delta F_\mathrm{ch} / \Delta F_\mathrm{line}$, and assume a rectangular frequency profile for both the line and the channel transmission for simplicity.

The SNR for $r=1$ after an integration time of $t$ is

\begin{equation}
\begin{split}
\mathrm{SNR} &= \frac{P_\mathrm{line}\sqrt{2t}}{\mathrm{NEP}_\mathrm{ph}}\\
\end{split}
\end{equation}

Here, $P_\mathrm{line}$ is the total power from the line absorbed by the MKID, and $\mathrm{NEP}_\mathrm{ph}$ is given by equation~\ref{eq:NEPph}.

If $r>1$, then $P_\mathrm{line}$ stays constant, but the NEP increases with $\sqrt{r}$ according to equation~\ref{eq:NEPph}, because $P_\mathrm{MKID}\propto r$ and $\Delta F\propto r$. In other words, the MKID receives more sky loading but the signal power from the line stays constant. Therefore the SNR drops. 

If $r<1$, then $P_\mathrm{line}$ decreases in proportion to $r$ because the channel receives only part of the spectral power of the line. At the same time the NEP decreases but with $\sqrt{r}$, so the net change in SNR is a decrease proportional to $\sqrt{r}$.

From these two arguments we can conclude that the SNR for a single channel is maximized for $r=1$. 

Now, the sensitivity loss for the $r<1$ case can be recovered if one places $1/r$ channels per line, but this would naturally require more detectors to cover a given instantaneous bandwidth. For DESHIMA we have taken a typical velocity width of $\Delta V \sim 600$ km/s for bright SMGs \cite{2013ARA&A..51..105C} to set the frequency resolution to $F/\Delta F = c/\Delta V \sim 500$. If the typical line width of the target population is known a priori, then the spectral resolution of the ISS can be optimized according to the intended applications.

\subsection*{Data availability} The datasets generated and analyzed during this study are available from the corresponding author on reasonable request. 

\subsection*{Code availability} The De:code software is distributed under the MIT license at \url{https://github.com/deshima-dev/decode}. 


\begin{thebibliography}{10}
\expandafter\ifx\csname url\endcsname\relax
  \def\url#1{\texttt{#1}}\fi
\expandafter\ifx\csname urlprefix\endcsname\relax\def\urlprefix{URL }\fi
\providecommand{\bibinfo}[2]{#2}
\providecommand{\eprint}[2][]{\url{#2}}

\bibitem{2019arXiv190304779G}
\bibinfo{author}{{Geach}, J.~E.} \emph{et~al.}
\newblock \bibinfo{title}{{The case for a 'sub-millimeter SDSS': a 3D map of
  galaxy evolution to z$\sim$10}}.
\newblock \emph{\bibinfo{journal}{arXiv e-prints}}  (\bibinfo{year}{2019}).
\newblock \eprint{1903.04779}.

\bibitem{2017arXiv170902389F}
\bibinfo{author}{{Farrah}, D.} \emph{et~al.}
\newblock \bibinfo{title}{{Review: Far-Infrared Instrumentation and Technology
  Development for the Next Decade}}.
\newblock \emph{\bibinfo{journal}{J. Astron. Telesc. Instrum. Syst.}}
  \textbf{\bibinfo{volume}{5}}, \bibinfo{pages}{020901} (\bibinfo{year}{2019}).

\bibitem{2016SPIE.9906E..26K}
\bibinfo{author}{Kawabe, R.} \emph{et~al.}
\newblock \bibinfo{title}{{New 50-m-class single-dish telescope: Large
  Submillimeter Telescope (LST)}}.
\newblock \emph{\bibinfo{journal}{Proc. SPIE}} \textbf{\bibinfo{volume}{9906}},
  \bibinfo{pages}{990626} (\bibinfo{year}{2016}).

\bibitem{2007ASPC..375...71E}
\bibinfo{author}{Erickson, N.}, \bibinfo{author}{Narayanan, G.},
  \bibinfo{author}{Goeller, R.} \& \bibinfo{author}{Grosslein, R.}
\newblock \bibinfo{title}{{An Ultra-Wideband Receiver and Spectrometer for
  74--110 GHz}}.
\newblock In \bibinfo{editor}{Baker, A.~J.}, \bibinfo{editor}{Glenn, J.},
  \bibinfo{editor}{Harris, A.~I.}, \bibinfo{editor}{Mangum, J.~G.} \&
  \bibinfo{editor}{Yun, M.~S.} (eds.) \emph{\bibinfo{booktitle}{From Z-Machines
  to ALMA: (Sub)Millimeter Spectroscopy of Galaxies}}, \bibinfo{pages}{71--81}
  (\bibinfo{year}{2007}).

\bibitem{2011ITTST...1..241S}
\bibinfo{author}{Stacey, G.~J.}
\newblock \bibinfo{title}{{THz Low Resolution Spectroscopy for Astronomy}}.
\newblock \emph{\bibinfo{journal}{IEEE Trans. THz Sci. Technol.}}
  \textbf{\bibinfo{volume}{1}}, \bibinfo{pages}{241--255}
  (\bibinfo{year}{2011}).

\bibitem{E1}
\bibinfo{author}{{Endo}, A.} \emph{et~al.}
\newblock \bibinfo{title}{{Wideband on-chip terahertz spectrometer based on a
  superconducting filterbank}}.
\newblock \emph{\bibinfo{journal}{arXiv e-prints}}  (\bibinfo{year}{2019}).
\newblock \eprint{1901.06934}.

\bibitem{2003Natur.425..817D}
\bibinfo{author}{{Day}, P.~K.}, \bibinfo{author}{{LeDuc}, H.~G.},
  \bibinfo{author}{{Mazin}, B.~A.}, \bibinfo{author}{{Vayonakis}, A.} \&
  \bibinfo{author}{{Zmuidzinas}, J.}
\newblock \bibinfo{title}{{A broadband superconducting detector suitable for
  use in large arrays}}.
\newblock \emph{\bibinfo{journal}{Nature}} \textbf{\bibinfo{volume}{425}},
  \bibinfo{pages}{817--821} (\bibinfo{year}{2003}).

\bibitem{2018Sci...362.1034D}
\bibinfo{author}{{D{\'\i}az-Santos}, T.} \emph{et~al.}
\newblock \bibinfo{title}{{The multiple merger assembly of a hyperluminous
  obscured quasar at redshift 4.6}}.
\newblock \emph{\bibinfo{journal}{Science}} \textbf{\bibinfo{volume}{362}},
  \bibinfo{pages}{1034--1036} (\bibinfo{year}{2018}).

\bibitem{Casey:2014gr}
\bibinfo{author}{Casey, C.~M.}, \bibinfo{author}{Narayanan, D.} \&
  \bibinfo{author}{Cooray, A.}
\newblock \bibinfo{title}{{Dusty star-forming galaxies at high redshift}}.
\newblock \emph{\bibinfo{journal}{Physics Reports}}
  \textbf{\bibinfo{volume}{541}}, \bibinfo{pages}{45--161}
  (\bibinfo{year}{2014}).

\bibitem{2010SPIE.7741E..0XG}
\bibinfo{author}{{Groppi}, C.} \emph{et~al.}
\newblock \bibinfo{title}{{Test and integration results from SuperCam: a
  64-pixel array receiver for the 350 GHz atmospheric window}}.
\newblock \emph{\bibinfo{journal}{Proc. SPIE}} \textbf{\bibinfo{volume}{7741}},
  \bibinfo{pages}{77410X} (\bibinfo{year}{2010}).

\bibitem{2012JLTP..167..341E}
\bibinfo{author}{Endo, A.} \emph{et~al.}
\newblock \bibinfo{title}{{Design of an Integrated Filterbank for DESHIMA:
  On-Chip Submillimeter Imaging Spectrograph Based on Superconducting
  Resonators}}.
\newblock \emph{\bibinfo{journal}{J. Low Temp. Phys.}}
  \textbf{\bibinfo{volume}{167}}, \bibinfo{pages}{341--346}
  (\bibinfo{year}{2012}).

\bibitem{Shirokoff:2012fx}
\bibinfo{author}{Shirokoff, E.} \emph{et~al.}
\newblock \bibinfo{title}{{MKID development for SuperSpec: an on-chip, mm-wave,
  filter-bank spectrometer}}.
\newblock \emph{\bibinfo{journal}{Proc. SPIE}} \textbf{\bibinfo{volume}{8452}},
  \bibinfo{pages}{84520R} (\bibinfo{year}{2012}).

\bibitem{Cataldo2018}
\bibinfo{author}{Cataldo, G.} \emph{et~al.}
\newblock \bibinfo{title}{Second-generation design of micro-spec: A
  medium-resolution, submillimeter-wavelength spectrometer-on-a-chip}.
\newblock \emph{\bibinfo{journal}{J. Low Temp. Phys.}}
  \textbf{\bibinfo{volume}{193}}, \bibinfo{pages}{923--930}
  (\bibinfo{year}{2018}).

\bibitem{2014SPIE.9150E..0JS}
\bibinfo{author}{{Sibthorpe}, B.} \& \bibinfo{author}{{Jellema}, W.}
\newblock \bibinfo{title}{{Relative performance of dispersive and
  non-dispersive far-infrared spectrometer instrument architectures}}.
\newblock \emph{\bibinfo{journal}{Proc. SPIE}} \textbf{\bibinfo{volume}{9153}},
  \bibinfo{pages}{91531W} (\bibinfo{year}{2014}).

\bibitem{2017ApPhL.110w3503B}
\bibinfo{author}{Bueno, J.} \emph{et~al.}
\newblock \bibinfo{title}{{Full characterisation of a background limited
  antenna coupled KID over an octave of bandwidth for THz radiation}}.
\newblock \emph{\bibinfo{journal}{Appl. Phys. Lett.}}
  \textbf{\bibinfo{volume}{110}}, \bibinfo{pages}{233503}
  (\bibinfo{year}{2017}).

\bibitem{OBrient:2013hc}
\bibinfo{author}{O'Brient, R.} \emph{et~al.}
\newblock \bibinfo{title}{{A dual-polarized broadband planar antenna and
  channelizing filter bank for millimeter wavelengths}}.
\newblock \emph{\bibinfo{journal}{Appl. Phys. Lett.}}
  \textbf{\bibinfo{volume}{102}}, \bibinfo{pages}{063506--5}
  (\bibinfo{year}{2013}).

\bibitem{2018NatAs...2..596B}
\bibinfo{author}{{Battersby}, C.} \emph{et~al.}
\newblock \bibinfo{title}{{The Origins Space Telescope}}.
\newblock \emph{\bibinfo{journal}{Nature Astronomy}}
  \textbf{\bibinfo{volume}{2}}, \bibinfo{pages}{596--599}
  (\bibinfo{year}{2018}).

\bibitem{2014NatCo...5E3130D}
\bibinfo{author}{{de Visser}, P.~J.}, \bibinfo{author}{{Baselmans}, J.~J.~A.},
  \bibinfo{author}{{Bueno}, J.}, \bibinfo{author}{{Llombart}, N.} \&
  \bibinfo{author}{{Klapwijk}, T.~M.}
\newblock \bibinfo{title}{{Fluctuations in the electron system of a
  superconductor exposed to a photon flux}}.
\newblock \emph{\bibinfo{journal}{Nature Communications}}
  \textbf{\bibinfo{volume}{5}}, \bibinfo{pages}{3130} (\bibinfo{year}{2014}).

\bibitem{Endo:2013ky}
\bibinfo{author}{Endo, A.} \emph{et~al.}
\newblock \bibinfo{title}{{On-chip filter bank spectroscopy at
  600{\textendash}700 GHz using NbTiN superconducting resonators}}.
\newblock \emph{\bibinfo{journal}{Appl. Phys. Lett.}}
  \textbf{\bibinfo{volume}{103}}, \bibinfo{pages}{032601}
  (\bibinfo{year}{2013}).

\bibitem{Wheeler:2016dr}
\bibinfo{author}{Wheeler, J.} \emph{et~al.}
\newblock \bibinfo{title}{{SuperSpec: development towards a full-scale filter
  bank}}.
\newblock \emph{\bibinfo{journal}{Proc. SPIE}} \textbf{\bibinfo{volume}{9914}},
  \bibinfo{pages}{99143K} (\bibinfo{year}{2016}).

\bibitem{2004SPIE.5489..763E}
\bibinfo{author}{{Ezawa}, H.}, \bibinfo{author}{{Kawabe}, R.},
  \bibinfo{author}{{Kohno}, K.} \& \bibinfo{author}{{Yamamoto}, S.}
\newblock \bibinfo{title}{{The Atacama Submillimeter Telescope Experiment
  (ASTE)}}.
\newblock \emph{\bibinfo{journal}{Proc. SPIE}} \textbf{\bibinfo{volume}{5489}},
  \bibinfo{pages}{763--772} (\bibinfo{year}{2004}).

\bibitem{2008ApJS..178..189W}
\bibinfo{author}{{Wilson}, C.~D.} \emph{et~al.}
\newblock \bibinfo{title}{{Luminous Infrared Galaxies with the Submillimeter
  Array. I. Survey Overview and the Central Gas to Dust Ratio}}.
\newblock \emph{\bibinfo{journal}{Astrophys. J., Suppl. Ser.}}
  \textbf{\bibinfo{volume}{178}}, \bibinfo{pages}{189--224}
  (\bibinfo{year}{2008}).

\bibitem{2016MNRAS.457.2139C}
\bibinfo{author}{{Coud{\'e}}, S.} \emph{et~al.}
\newblock \bibinfo{title}{{The JCMT Gould Belt Survey: the effect of molecular
  contamination in SCUBA-2 observations of Orion A}}.
\newblock \emph{\bibinfo{journal}{Mon. Notices Royal Astron. Soc.}}
  \textbf{\bibinfo{volume}{457}}, \bibinfo{pages}{2139--2150}
  (\bibinfo{year}{2016}).

\bibitem{2001A&A...373..853D}
\bibinfo{author}{{Dumke}, M.}, \bibinfo{author}{{Nieten}, C.},
  \bibinfo{author}{{Thuma}, G.}, \bibinfo{author}{{Wielebinski}, R.} \&
  \bibinfo{author}{{Walsh}, W.}
\newblock \bibinfo{title}{{Warm gas in central regions of nearby galaxies.
  Extended mapping of CO(3-2) emission}}.
\newblock \emph{\bibinfo{journal}{Astron. Astrophys.}}
  \textbf{\bibinfo{volume}{373}}, \bibinfo{pages}{853--880}
  (\bibinfo{year}{2001}).

\bibitem{Bueno2018}
\bibinfo{author}{{Yurduseven}, O.} \emph{et~al.}
\newblock \bibinfo{title}{Incoherent detection of orthogonal polarizations via
  an antenna coupled {MKID}: Experimental validation at 1.55 {THz}}.
\newblock \emph{\bibinfo{journal}{IEEE Trans. Terahertz Sci. Technol.}}
  \textbf{\bibinfo{volume}{8}}, \bibinfo{pages}{736--745}
  (\bibinfo{year}{2018}).

\bibitem{Ito:2018bb}
\bibinfo{author}{Ito, T.} \emph{et~al.}
\newblock \bibinfo{title}{{The new heterodyne receiver system for the ASTE
  radio telescope: three-cartridge cryostat with two cartridge-type
  superconducting receivers}}.
\newblock \emph{\bibinfo{journal}{Proc. SPIE}}
  \textbf{\bibinfo{volume}{10708}}, \bibinfo{pages}{107082V}
  (\bibinfo{year}{2018}).

\bibitem{2018Natur.562..229W}
\bibinfo{author}{{Wisotzki}, L.} \emph{et~al.}
\newblock \bibinfo{title}{{Nearly all the sky is covered by Lyman-{$\alpha$}
  emission around high-redshift galaxies}}.
\newblock \emph{\bibinfo{journal}{Nature}} \textbf{\bibinfo{volume}{562}},
  \bibinfo{pages}{229--232} (\bibinfo{year}{2018}).

\bibitem{2018SPIE10700E..5XP}
\bibinfo{author}{{Parshley}, S.~C.} \emph{et~al.}
\newblock \bibinfo{title}{{CCAT-prime: a novel telescope for sub-millimeter
  astronomy}}.
\newblock \emph{\bibinfo{journal}{Proc. SPIE}}
  \textbf{\bibinfo{volume}{10700}}, \bibinfo{pages}{107005X}
  (\bibinfo{year}{2018}).

\bibitem{6111333}
\bibinfo{author}{Mahieu, S.} \emph{et~al.}
\newblock \bibinfo{title}{The alma band-7 cartridge}.
\newblock \emph{\bibinfo{journal}{IEEE Trans. Terahertz Sci. Technol.}}
  \textbf{\bibinfo{volume}{2}}, \bibinfo{pages}{29--39} (\bibinfo{year}{2012}).

\bibitem{2001ITAP...49.1683P}
\bibinfo{author}{{Pardo}, J.~R.}, \bibinfo{author}{{Cernicharo}, J.} \&
  \bibinfo{author}{{Serabyn}, E.}
\newblock \bibinfo{title}{{Atmospheric transmission at microwaves (ATM): an
  improved model for millimeter/submillimeter applications}}.
\newblock \emph{\bibinfo{journal}{IEEE Trans. Antennas Propag.}}
  \textbf{\bibinfo{volume}{49}}, \bibinfo{pages}{1683--1694}
  (\bibinfo{year}{2001}).

\bibitem{2013A&A...552A.104N}
\bibinfo{author}{{Nikolic}, B.}, \bibinfo{author}{{Bolton}, R.~C.},
  \bibinfo{author}{{Graves}, S.~F.}, \bibinfo{author}{{Hills}, R.~E.} \&
  \bibinfo{author}{{Richer}, J.~S.}
\newblock \bibinfo{title}{{Phase correction for ALMA with 183 GHz water vapour
  radiometers}}.
\newblock \emph{\bibinfo{journal}{Astron. Astrophys.}}
  \textbf{\bibinfo{volume}{552}}, \bibinfo{pages}{A104} (\bibinfo{year}{2013}).

\bibitem{Butler2012}
\bibinfo{author}{Butler, B.}
\newblock \bibinfo{title}{Flux density models for solar system bodies in casa}.
\newblock \emph{\bibinfo{journal}{ALMA Memo}} \textbf{\bibinfo{volume}{594}}
  (\bibinfo{year}{2012}).

\bibitem{ToolsOfRadioAstronomy}
\bibinfo{author}{{Wilson}, T.~L.}, \bibinfo{author}{{Rohlfs}, K.} \&
  \bibinfo{author}{{H{\"u}ttemeister}, S.}
\newblock \emph{\bibinfo{title}{{Tools of Radio Astronomy}}}
  (\bibinfo{year}{2009}).

\bibitem{decode}
\bibinfo{title}{{DESHIMA Code for data analysis; De:code}}.
\newblock \bibinfo{note}{\url{https://github.com/deshima-dev/decode}}.

\bibitem{Zemax}
\bibinfo{title}{{Zemax LLC, Zemax}}.
\newblock \bibinfo{note}{\url{https://www.zemax.com}}.

\bibitem{Rantwijk2016}
\bibinfo{author}{van Rantwijk, J.}, \bibinfo{author}{Grim, M.},
  \bibinfo{author}{van Loon, D.} \& \bibinfo{author}{Yates, S. J.~C.}
\newblock \bibinfo{title}{{Multiplexed Readout for 1000-Pixel Arrays of
  Microwave Kinetic Inductance Detectors}}.
\newblock \emph{\bibinfo{journal}{IEEE Trans. Microwave Theory Techn.}}
  \textbf{\bibinfo{volume}{64}}, \bibinfo{pages}{1876--1883}
  (\bibinfo{year}{2016}).

\bibitem{2011ApJS..193...17P}
\bibinfo{author}{{Patel}, N.~A.} \emph{et~al.}
\newblock \bibinfo{title}{{An Interferometric Spectral-line Survey of IRC+10216
  in the 345 GHz Band}}.
\newblock \emph{\bibinfo{journal}{Astrophys. J., Suppl. Ser.}}
  \textbf{\bibinfo{volume}{193}}, \bibinfo{pages}{17} (\bibinfo{year}{2011}).

\bibitem{1992A&A...266..365W}
\bibinfo{author}{{Williams}, P.~G.} \& \bibinfo{author}{{White}, G.~J.}
\newblock \bibinfo{title}{{Sub-millimetre molecular lines in the circumstellar
  envelope IRC+10216}}.
\newblock \emph{\bibinfo{journal}{Astron. Astrophys.}}
  \textbf{\bibinfo{volume}{266}}, \bibinfo{pages}{365--376}
  (\bibinfo{year}{1992}).

\bibitem{1994ApJS...95..503W}
\bibinfo{author}{{Wang}, Y.}, \bibinfo{author}{{Jaffe}, D.~T.},
  \bibinfo{author}{{Graf}, U.~U.} \& \bibinfo{author}{{Evans}, N.~J., II}.
\newblock \bibinfo{title}{{Single-sideband calibration for CO, (13)CO, HCN, and
  CS lines near 345 GHz}}.
\newblock \emph{\bibinfo{journal}{Astrophys. J., Suppl. Ser.}}
  \textbf{\bibinfo{volume}{95}}, \bibinfo{pages}{503--515}
  (\bibinfo{year}{1994}).

\bibitem{2006AJ....131.1163S}
\bibinfo{author}{{Skrutskie}, M.~F.} \emph{et~al.}
\newblock \bibinfo{title}{{The Two Micron All Sky Survey (2MASS)}}.
\newblock \emph{\bibinfo{journal}{Astrophys. J.}}
  \textbf{\bibinfo{volume}{131}}, \bibinfo{pages}{1163--1183}
  (\bibinfo{year}{2006}).

\bibitem{2004A&A...427...45B}
\bibinfo{author}{{Bayet}, E.}, \bibinfo{author}{{Gerin}, M.},
  \bibinfo{author}{{Phillips}, T.~G.} \& \bibinfo{author}{{Contursi}, A.}
\newblock \bibinfo{title}{{The submillimeter C and CO lines in Henize 2-10 and
  NGC 253}}.
\newblock \emph{\bibinfo{journal}{Astron. Astrophys.}}
  \textbf{\bibinfo{volume}{427}}, \bibinfo{pages}{45--59}
  (\bibinfo{year}{2004}).

\bibitem{2014SuScT..27e5012G}
\bibinfo{author}{Guruswamy, T.}, \bibinfo{author}{Goldie, D.~J.} \&
  \bibinfo{author}{Withington, S.}
\newblock \bibinfo{title}{{Quasiparticle generation efficiency in
  superconducting thin films}}.
\newblock \emph{\bibinfo{journal}{Supercond. Sci. Technol.}}
  \textbf{\bibinfo{volume}{27}}, \bibinfo{pages}{055012}
  (\bibinfo{year}{2014}).

\bibitem{2013ARA&A..51..105C}
\bibinfo{author}{{Carilli}, C.~L.} \& \bibinfo{author}{{Walter}, F.}
\newblock \bibinfo{title}{{Cool Gas in High-Redshift Galaxies}}.
\newblock \emph{\bibinfo{journal}{Annual Review of Astronomy and Astrophysics}}
  \textbf{\bibinfo{volume}{51}}, \bibinfo{pages}{105--161}
  (\bibinfo{year}{2013}).
\newblock \eprint{1301.0371}.

\end{thebibliography}

\subsection*{Correspondence and requests for materials}
Correspondence and requests for materials should be addressed to A.E.

\section*{Acknowledgements}
We thank Toshihiko Kobiki, Tetsuya Ito, Masumi Yamada, Motoi Saito, Javier Aguilera, and Javier Zenteno of NAOJ for their support at ASTE. We thank Ricardo Jara, Lorena Toro Galv\'{e}z, Mika Konuma of NAOJ for their support in the transportation of the equipment to ASTE. We thank Tetsuhiro Minamidani for hosting a go/no-go review of the campaign, and all committee members who provided invaluable feedback.  We thank Klaas Keizer of SRON for the precise mechanical work on the cryostat. We thank Peter Paul Kooijman and Henk Hoevers of SRON for coordinating the delivery of the cryogenic hardware. We thank the staff of The University of Tokyo Atacama Observatory facility for their hospitality. We thank the staff of Kavli Nanolab Delft for their support in the microfabrication of the ISS chip. We thank the staff of Else Kooi Labratory for supporting the measurements in the cryolab at TU Delft. We thank Doreen Wernicke and Josef Baumgartner of Entropy Cryogenics for the support in operating the cryostat at ASTE.  Finally, we thank Jos\'{e} Pinto for his kindness to donate a piece of copper wire with a diameter in the range of 1.00--1.05 mm from his jewelry shop in San Pedro de Atacama, so that we could align the cryogenic thermal mechanical structure on site. This research was supported by the Netherlands Organization for Scientific Research NWO (Vidi grant No. 639.042.423, NWO Medium Investment grant No. 614.061.611 DESHIMA), the European Research Counsel ERC (ERC-CoG-2014 - Proposal n$^\circ$ 648135 MOSAIC), the Japan Society for the Promotion of Science JSPS (KAKENHI Grant Numbers JP25247019 and JP17H06130), NAOJ ALMA Scientific Research Grant Number 2018-09B, and the Grant for Joint Research Program of the Institute of Low Temperature Science, Hokkaido University. P.J. de V. is supported by the NWO (Veni Grant 639.041.750). T.M.K. is supported by the ERC Advanced Grant No. 339306 (METIQUM) and the Russian Science Foundation (Grant No. 17-72-30036). N.L. is supported by ERC (Starting Grant No. 639749). J.S. and M.N. are supported by the JSPS Program for Advancing Strategic International Networks to Accelerate the Circulation of Talented Researchers (Program No. R2804). T.J.L.C.B. was supported by the European Union Seventh Framework Programme (FP7/2007--2013, FP7/2007--2011) under grant agreement No. 607254. The ASTE telescope is operated by National Astronomical Observatory of Japan (NAOJ).

\section*{Author contributions}

A.E. initiated the DESHIMA project as an MKID-based redshift machine. J.J.A.B. invented the concept of the ISS. P.P. van der W., Y.T., Kohno, and R.K. articulated further astronomical usage of ISS spectrometers. A.E. designed the ISS filterbank. O.Y. designed the double-slot antenna. A.P.L. explained the chip performance with precise electromagnetic simulations. D.J.T. and V.M. fabricated the chip. D.J.T. and T.M.K. provided the NbTiN thin film. J.B. measured the optical efficiency of the chip. P.J. de V. provided insight on the quasiparticle physics. S.J.C.Y. designed the cold optics, measured the instrument beam pattern, and did a post analysis to explain the beam pattern and efficiency measured on ASTE. J.J.A.B. and S.J.C.Y. made the conceptual design of the cryogenic setup, and R.H. made the mechanical designs. J.J.A.B. developed the readout electronics. Karatsu measured the sensitivity and frequency-response of the instrument. M.N. and J.S. contributed to these measurements. J.S. developed a database for managing the acquired data. S.B., O.Y. and N.L. designed the warm optics. T.O., Takekoshi, K.O., and Y.F. designed and tested the warm optics, the room temperature calibration chopper, and the DESHIMA-ASTE hardware interface. A.K. and K.F. manufactured the warm optics, and S.N. measured its surface accuracy. Karatsu, Y.T. and J.M. developed the DESHIMA local controller. D.J.T. and T.O. were responsible for the logistics in the transportation of the equipment to ASTE. T.O. led the installation of DESHIMA on ASTE, done by T.O., Takekoshi, Karatsu, D.J.T., R.H., and A.E.. R.H. and Karatsu were responsible for the re-integration of the DESHIMA hardware on the ASTE site. Karatsu and T.O. realized remote control of DESHIMA on ASTE. Takekoshi aligned the warm optics using the scheme he developed. Ishii, A.T., Y.T., Karatsu, Takekoshi, T.U., T.I., K.C., and K.S. defined the data structure. A.T. and T.I. developed the De:code software. Y.T. led the astronomical observations and selected the target objects. Observations were conducted from the TAO facility in San Pedro de Atacama and from NAOJ by Y.T., K.S., T.I., A.T., Takekoshi, T.O., Karatsu, K.C., Y.Y., T.J.L.C.B., Ishii, T.U., and A.E.. Y.T. developed the on-sky chopping scheme. Takekoshi led the dismounting of DESHIMA, done by Takekoshi, Karatsu, M.N., K.F., and A.E. The following authors autonomously analyzed the on-telescope data and wrote the corresponding sections of this paper: K.S. (Mars), Takekoshi (sky dip calibration, in collaboration with J.S. and Karatsu), Tsukagoshi (VV~114, IRC+10216), Ikarashi (Orion, NGC~253). A.E. led the writing of the paper, and all authors have contributed to improving the quality. Project management: S.A. managed the ASTE telescope; J.J.A.B. managed the development of the instrument hardware; T.O. managed the development of the warm optics and chopper, as well as the scheme and hardware for installing DESHIMA on ASTE; Y.T. managed the astronomical commissioning and software development; A.E. managed the DESHIMA project on the top level.


\newpage
\clearpage
\onecolumngrid
\setcounter{figure}{0}
\setcounter{page}{1}
\setcounter{equation}{0}
\renewcommand{\theequation}{S\arabic{equation}}
\renewcommand{\thefigure}{S\arabic{figure}}
\renewcommand{\figurename}{\textbf{Supplementary Fig.}}
\renewcommand{\thesection}{Supplementary Note \arabic{section}}
\renewcommand{\thetable}{S\arabic{table}}
\renewcommand{\tablename}{\textbf{Supplementary Table}}

\noindent\textbf{\textsf{Supplementary Information For: \\First light demonstration of the integrated superconducting spectrometer}}

\vspace{10pt}

\noindent
\textsf{Akira~Endo, Kenichi~Karatsu, Yoichi~Tamura, Tai~Oshima,
Akio~Taniguchi, Tatsuya~Takekoshi, Shinâichiro~Asayama, Tom~J.L.C.~Bakx,
Sjoerd~Bosma, Juan~Bueno, Kah~Wuy~Chin, Yasunori~Fujii, Kazuyuki~Fujita,
Robert~Huiting, Soh~Ikarashi, Tsuyoshi~Ishida, Shun~Ishii, Ryohei~Kawabe,
Teun~M.~Klapwijk, Kotaro~Kohno, Akira~Kouchi, Nuria~Llombart,
Jun~Maekawa, Vignesh~Murugesan, Shunichi~Nakatsubo, Masato~Naruse,
Kazushige~Ohtawara, Alejandro~Pascual~Laguna, Junya~Suzuki, Koyo~Suzuki,
David~J.~Thoen, Takashi~Tsukagoshi, Tetsutaro~Ueda, Pieter~J.~de~Visser,
Paul~P.~van~der~Werf, Stephen~J.C.~Yates, Yuki~Yoshimura, Ozan~Yurduseven, and Jochem~J.A.~Baselmans}

\begin{figure*}[htbp]
 \includegraphics[width=0.6\textwidth]{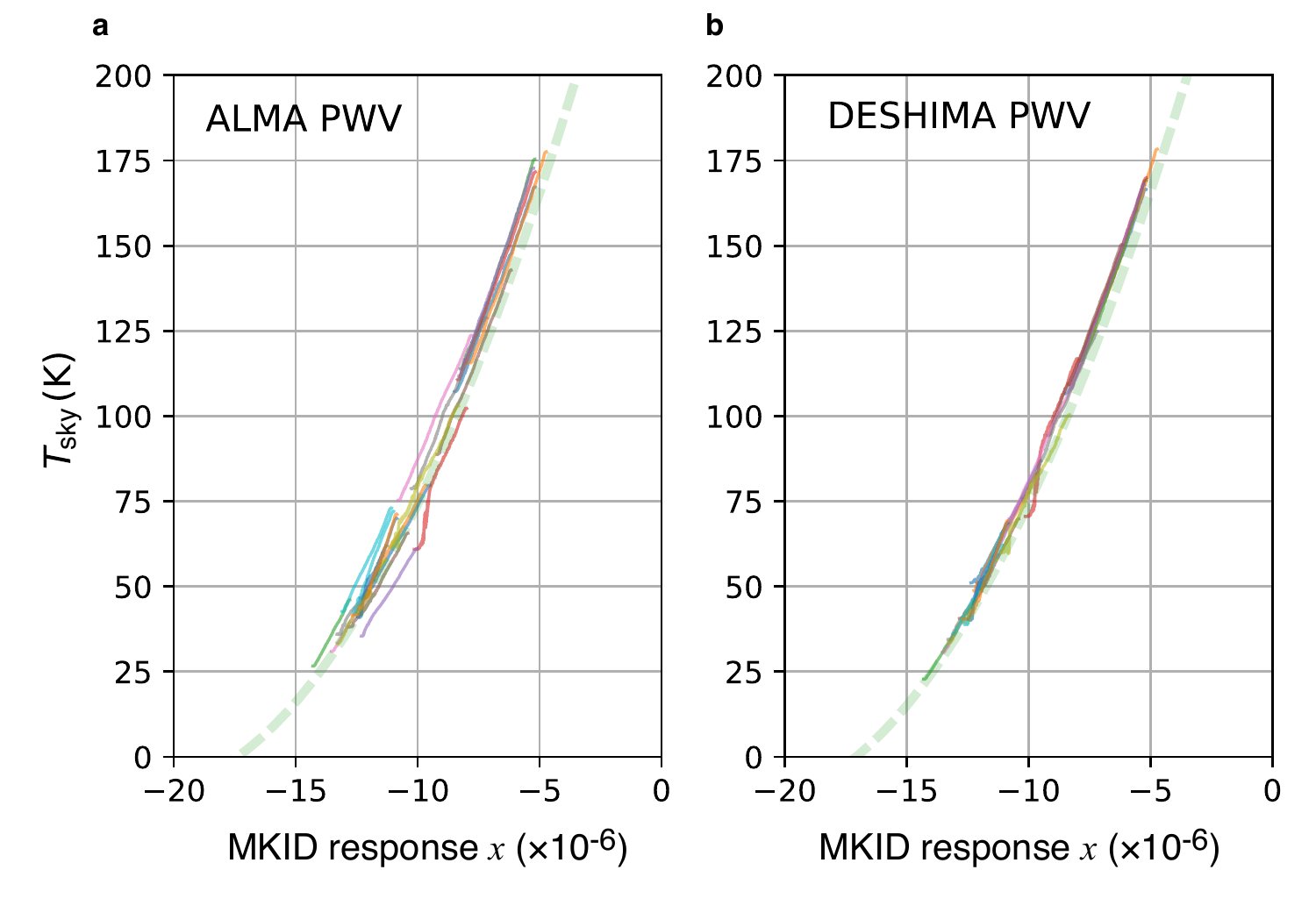}
 \linespread{1.0}\selectfont
 \caption{\label{fig:calibration} 
  \textbf{Conversion from the relative MKID frequency response $x$ to sky brightness temperature $T_\mathrm{sky}$.}  
   \textbf{a}, Sky temperature $T_\mathrm{sky}$ as a function of relative MKID frequency shift $x$ for a representative filterbank channel. 
   $x$ is defined relative to when an ambient temperature black body is loading the beam coming out of the cryostat: $x\equiv \{f(T_\mathrm{sky})-f(T_\mathrm{amb} )\}/f(T_\mathrm{amb})$.
   The analysis is based on time-dependent PWV values taken directly from the ALMA radiometer\citep{2013A&A...552A.104N} without any correction. Each solid curve is taken with one skydip measurement. The dashed curve is a square-law fit to all measurements. \textbf{b}, Improved calibration model, after an iterative parameter fit which takes the PWV as a fitting parameter that is common to all filterbank channels.
 }%
 \end{figure*}
 \begin{figure*}[htbp]
 \includegraphics[width=\textwidth]{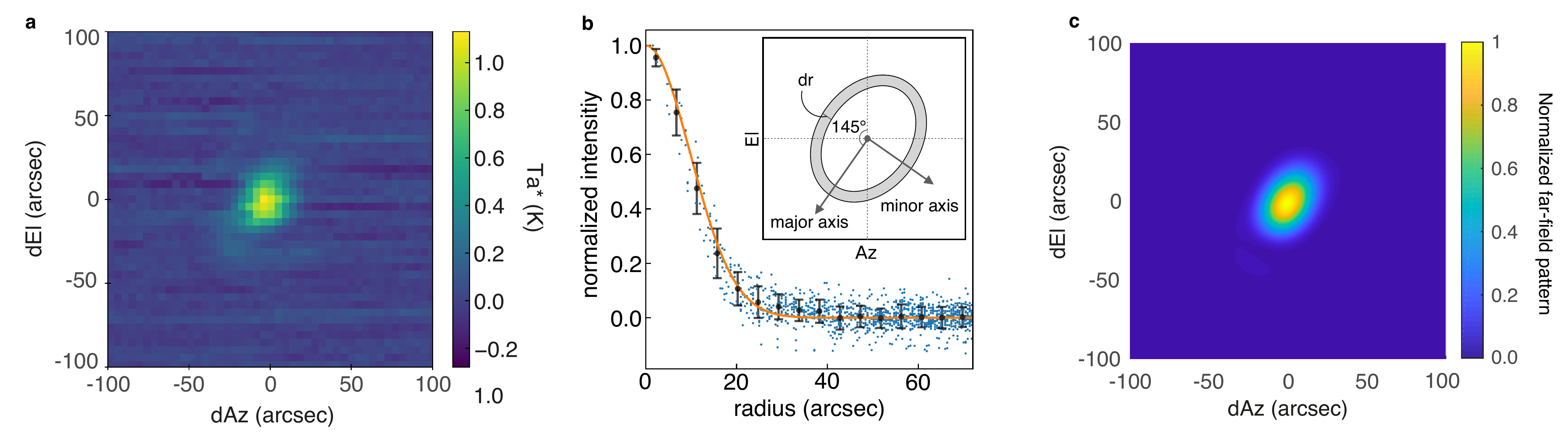}%
 \linespread{1.0}\selectfont
 \caption{\label{fig:beampattern}
 \textbf{The beam pattern of ASTE/DESHIMA.}
  \textbf{a}, The DESHIMA 350~GHz image of Mars. The response of all spectral channels are added together. 
  The color scale represents antenna temperature.
  \textbf{b}, The radial profile of the 350 GHz image of Mars with an apparent diameter of 3.99$''$. 
  Since the beam shape is elliptical, the normalized intensities (blue dots) are measured with elliptical annuli as a function of their minor radii. The radius of each pixel in the image is defined as minor axis of an ellipse (corresponding to the beam shape of panel a) which crosses it, and the blue data points indicate the normalized intensities at each pixel in the image. The black points and error bars show the mean and standard deviation of the normalized intensity at a radial bin in steps of  $dr=4.5''$ (inset). After correcting for the ellipticity, the main beam is well described by a Gaussian with a FWHM of 22.8 $\pm$ 3.1 arcseconds (orange line).
  \textbf{c}, Normalized far-field beam pattern, simulated assuming a 4$^\circ$ offset from the ideal optical axis at the lens-antenna on chip.
}%
 \end{figure*}

\newpage

\section{Main beam size and efficiency}
\label{SI:efficiency}

We measure the main-beam shape and efficiency on a Mars image where Mars cannot be considered as a point source (i.e., $\Omega_\mathrm{source} \ll \Omega_\mathrm{MB}$, where the former and latter are the solid angles of the source and the main beam, respectively). The Mars image is produced by putting time-stream data, which are already flux-calibrated by the standard chopper-wheel method, into 2-dimensional pixels (Supplementary Fig.~\ref{fig:beampattern}a). Then we fit a 2-dimensional Gaussian to this continuum map to measure the intrinsic beam size and amplitude.  We obtain a peak antenna temperature of $T_\mathrm{a}^{\ast} = 1.06 \pm 0.1$~K and the half-power beam width (HPBW) of 31.4$''$ $\pm$ 2.8$''$ by 22.8$''$ $\pm$ 3.1$''$  with a position angle of 145 degrees. The beam radial profile is presented in Supplementary Fig.~\ref{fig:beampattern}b. 
The antenna temperature is related to the main beam efficiency as the equation
\begin{equation}
    T_\mathrm{a}^\ast = \frac{\int_\mathrm{source} P_n(\theta, \phi)\,d\Omega}{\int_{4\pi}P_n(\theta, \phi)d\Omega}T_\mathrm{brightness} = \eta_\mathrm{MB} \frac{\Omega_\mathrm{source}}{\Omega_\mathrm{MB}} T_\mathrm{brightness}\ ,
\end{equation}
where $P_n$ is the power pattern, $\Omega_\mathrm{MB}$ and $\Omega_\mathrm{source}$ are the solid angle of main beam and the source, respectively, and $\eta_\mathrm{MB}$ is the main beam efficiency.
The solid angle of an elliptical main beam is expressed as
$\Omega_\mathrm{MB} = \frac{\pi \theta_\mathrm{major}\theta_\mathrm{minor}}{4 \ln{2}}$, 
where $\theta_\mathrm{major}$ and $\theta_\mathrm{minor}$ are the FWHMs of the main beam measured along the major and minor axes, respectively. We regard the brightness distribution of Mars as a disk with a uniform intensity of $T_\mathrm{brightness} = 210$~K and an apparent diameter of 3.99$''$, and obtain $\eta_\mathrm{MB} = 0.34 \pm 0.03$ at 350~GHz.
\section{Jack-knife estimation of the noise level}\label{SI:jack-knife}

Since each MKID is an independent detector, the noise level of each ISS channel is determined by applying an iterative integration method with random sign inversion (the jack-knife method). 
In this method we apply the following steps to the time-stream data of each channel, independently.
First, we subtract the time-integrated signal of the source from the entire time-stream data. 
Then we divide this source-subtracted time-stream data into blocks that contain one cycle of the telescope nodding, and randomly invert the sign of the signal cotained in each block.
The time-integration of this randomized set of data yields a single estimation of the source-subtracted intensity of each channel. 
We repeat this process for 100 times, and use the standard deviation as the estimation for the noise level at that channel. The channel-dependent error of the broadband spectra of VV~114 and IRC+10216 as shown in Fig. \ref{fig:concept}b and Fig. \ref{fig:NEFD}a, as well as the NEFDs shown in Fig. \ref{fig:NEFD}c, are determined using this method.

\end{document}